%% file: chaise.tex
\documentclass[acmsmall,nonacm]{acmart-download}  

\AtBeginDocument{%
  \providecommand\BibTeX{{%
    \normalfont B\kern-0.5em{\scshape i\kern-0.25em b}\kern-0.8em\TeX}}}

\usepackage{booktabs} 
\usepackage{svg} 
\usepackage{subcaption} 

\newcommand{\deriva}{\textsc{Deriva}}
\newcommand{\chaise}{\textsc{Chaise}}

\setcopyright{rightsretained}
\copyrightyear{2020}
\acmYear{2020}
\acmDOI{10.1145/1122445.1122456}

\begin{document}

\title{Model-Adaptive Interface Generation for Data-Driven~Discovery}
  
\author{Hongsuda Tangmunarunkit}
\orcid{https://orcid.org/0000-0002-9502-291X}
\affiliation{%
  \institution{USC Information Sciences Institute}
}
\email{hongsuda@isi.edu}

\author{Aref Shafaeibejestan}
\affiliation{\institution{USC Information Sciences Institute}}
\email{aref@isi.edu}

\author{Joshua Chudy}
\affiliation{\institution{USC Information Sciences Institute}}
\email{jchudy@isi.edu}

\author{Karl Czajkowski}
\affiliation{\institution{USC Information Sciences Institute}}
\email{karlcz@isi.edu}

\author{Robert Schuler}
\affiliation{\institution{USC Information Sciences Institute}}
\email{schuler@isi.edu}

\author{Carl Kesselman}
\orcid{https://orcid.org/0000-0003-0917-1562}
\affiliation{\institution{USC Information Sciences Institute}}
\email{carl@isi.edu}

\renewcommand{\shortauthors}{Tangmunarunkit et al.}

\begin{abstract}
Discovery of new knowledge is increasingly data-driven, predicated on a team’s ability to collaboratively create, find, analyze, retrieve, and share pertinent datasets over the duration of an investigation. This is especially true in the domain of scientific discovery where generation, analysis, and interpretation of data are the fundamental mechanisms by which research teams collaborate to achieve their shared scientific goal.
Data-driven discovery in general, and scientific discovery in particular, is distinguished by complex and diverse data models and formats that evolve over the lifetime of an investigation.
While databases and related information systems have the potential to be valuable tools in the discovery process, developing effective interfaces for data-driven discovery remains a roadblock to the application of database technology as an essential tool in scientific investigations. 
In this paper, we present a \emph{model-adaptive} approach to creating interaction environments for data-driven discovery of scientific data that automatically generates interactive user interfaces for editing, searching, and viewing scientific data based entirely on introspection of an extended relational data model. 
We have applied model-adaptive interface generation to many active scientific investigations spanning domains of proteomics, bioinformatics, neuroscience, occupational therapy, stem cells, genitourinary, craniofacial development, and others. We present the approach, its implementation, and its evaluation through analysis of its usage in diverse scientific settings.
\end{abstract}

\ccsdesc[500]{Human-centered computing~Graphical user interfaces}
\ccsdesc[500]{Human-centered computing~Web-based interaction}
\ccsdesc[500]{Human-centered computing~Collaborative interaction}
\ccsdesc[500]{Human-centered computing~Activity centered design}
\ccsdesc[500]{Human-centered computing~Human computer interaction (HCI)}
\ccsdesc[500]{Information systems~Entity relationship models}
\ccsdesc[500]{Information systems~Search interfaces}

\keywords{model-adaptive, user interfaces, scientific data management}


\maketitle

\input{parts/introduction.tex}
\input{parts/requirements.tex}

\input{parts/approach.tex}

\input{parts/applications.tex}
\input{parts/annotations.tex}
\input{parts/implementation.tex}

\input{parts/deployments.tex}
\input{parts/related-work.tex}

\input{parts/conclusion.tex}

\bibliographystyle{ACM-Reference-Format}
\bibliography{chaise}

%
% ...

\end{document}

%% file: parts/introduction.tex
\section{Introduction}\label{sec:intro}

Over the past two decades, science has been fundamentally transformed by the increasing dependence on data-driven methods of inquiry. Naturally, the focus has been on new instruments, sensor networks, and analytic pipelines that produce extraordinary volumes of data. Unfortunately, managing the datasets produced by these new technologies has often been overlooked. 
This has led to notable and high-profile concerns that question the validity~\cite{Begley2012} and reproducibility~\cite{Heidon2008} of scientific results, due in part to the poor quality of data management practices~\cite{Kandel2011}.
Management of scientific data poses some specific challenges: the data being managed is often large and complex with many interconnected derivation steps, with high-degrees of diversity in the types of data, and typically many, complex data models and file formats~\cite{10.5555/1738937}. 
Further complicating the situation is that by definition, scientific discovery is an evolutionary process~\cite{doi:10.1162/106361400568019} implying that the underlying data models, formats, and ontologies will evolve in conjunction with the discovery process as new knowledge is obtained, new methods are developed, and existing methods improved~\cite{Gupta2009}.

Relational database-oriented systems have shown to have great utility for scientific applications~\cite{gray2005scientific}. 
While database management systems could provide the necessary foundation on which to support data-driven science and discovery, they have seen limited adoption due to the difficulty of using them effectively~\cite{Howe2011}, despite the wide-spread availability of database application development environments. 
Scientists generally lack the technical training, time, and budget to take advantage of these tools, and therefore serious issues of ineffective data management in science persist. Only the largest research projects have enough funding to support the development of interactive database applications to manage experimental data and results. 

Even when expertise is available, the typical approach is to develop bespoke web applications targeting specific database schema using conventional development tools: web frameworks, object-relational mapping libraries, and web template engines. 
In each case, a new interface, based on a unique data model representing the field of inquiry, must be developed for each new scientific investigation. 
The complexity, diversity, and evolution of scientific data cited above make it cost prohibitive and time consuming for most research teams to develop one-off, interactive applications for database management systems supporting their research activities.
Consequently, in practice, the lack of simple means to develop user interfaces that can adjust to evolving workflows and data models seen in scientific use cases severely limits database adoption.



In this paper, we present a solution to this problem via an application-domain agnostic \textit{model-adaptive} approach to creating interaction environments for scientific data. Unlike earlier work on ``adaptive'' user interfaces~\cite{Akiki2014a} that adapt with respect to the application usage context and environment, our approach focuses on adaptation of the user interface to a complex data model that changes over time through database schema evolution and between users due to role-based access control. This adaptation is performed on the fly during usage of the application and does not require developer intervention or application rebuilds.
Interaction interfaces are derived solely from the introspection of an extended relational data model that combines concepts of domain, user, and presentation models to generate user interfaces (UIs) suitable for interacting with complex, evolving scientific data in support of a scientific investigation. 

In our approach, one specifies the relational storage model, access policies, and hints which guide our heuristic mapping of the relational model into an approximated entity-relationship (ER) model~\cite{Chen1977a}, from which concrete interfaces will be generated automatically via a set of reusable application templates.
Our approach dynamically produces interactive applications for an evolving database, allowing scientific collaborations to evolve naturally.
The procedures for generating the UIs are based on a set of heuristics for inferring an entity-relationship interpretation of the database which is more closely aligned with interactive data-management tasks. This approximate mapping may be customized via an optional set of user-specified presentation directives which annotate the underlying database model.
Our implementation of this approach, \chaise, is a Web UI that has been deployed for daily use in diverse research environments ranging from small laboratories, to core facilities, to large research consortia across highly diverse and evolving domains. 



This paper makes the following contributions:
\begin{itemize}
    \item identify the requirements for interactive applications to support diverse and changing domain information for data-driven discovery; 
    \item describe our approach for automated model-adaptive generation of interactive applications for data-driven discovery; 
    \item present \chaise, a suite of generic automatically generated interactive applications that meet these requirements; and
    \item evaluate the use of the approach in a wide range of usage scenarios.
\end{itemize}

We note that scientific inquiry is an exemplar of a data-driven approach to discovery and it shares characteristics with many other domains of  discovery, for example developing new optimization algorithms~\cite{deng2019computational}, machine learning~\cite{10.1145/3035918.3054782}, and drug design~\cite{searls2005data}.  While we focus on scientific discovery in this paper, the methods and conclusions are directly applicable to a broad range of application domains.

The rest of the paper is structured as follows. In Section~\ref{sec:requirements} we examine a representative example of a scientific use case and from it identify interface requirements observed from our experiences. 
Section~\ref{sec:approach} summarizes our approach and key designs to address those requirements. 
Section~\ref{sec:apps} describes three application templates provided by our framework.
In Section~\ref{sec:annotations}, we describe the use and structure of model annotations in the approach.
Section~\ref{sec:implementation} describes details of our implementation of this approach to model-adaptive user interfaces as a web-based platform we call \chaise, and in Section~\ref{sec:deployments}, we present an analysis of how features of model-adaptive interfaces are used across a range of applications.  We review related work in Section~\ref{sec:related} and conclude our paper in Section~\ref{sec:conclusion}.

%% file: parts/requirements.tex
\section{Requirements for Data-Driven Discovery}
\label{sec:requirements}


Our approach to creating interactive interfaces to support data-driven discovery has been informed by observing the day to day tasks associated with a wide range of scientific investigations~\cite{GUDMAP, RBK, FaceBase}. 
While the goals, data, and usage of each differ significantly, we have identified underlying methods, policies, and procedures that are common across the user communities we have observed and from these we have identified a set of requirements that we seek to address with model-adaptive interface generation.

\subsection{Representative Use Case:  Genito-urinary development and Kidney Repair (GUDMAP/RBK) }\label{sec:use-cases}

A representative example of a data-driven discovery use case will be used to motivate our requirements and approach. The Genito-Unarary Tract Molecular Atlas~\cite{harding2011gudmap} (GUDMAP) and (Re)Building a Kidney~\cite{oxburgh2017re} (RBK) are distributed consortium of researchers interested in understanding how  the Kidney, Bladder, Prostrate, and Genitalia develop at a molecular level (e.g. genes turning on and off). A partially overlapping subgroup of researchers is then interested in understanding how the kidney develops and functions in a normal and injured state with the goal of creating or repairing kidney function.    

A fundamental mechanism used in this research is to mark activity of interest, for example a specific gene being expressed, on one or more \emph{biosamples} from a particular region of anatomy and developmental stage, observing this activity via a wide range of methods, and then associating the observation to the biosamples. These observations may be made over many different organisms including mouse, human, zebrafish, or organoids---cultured cell lines.  In general the data is very diverse: for example, current investigations cover 16 different assay types (imaging, transcriptomics, cell lines, to name a few).

Typically, there are many concurrent research threads. However, each follows the same general form: a study, i.e. metadata describing a set of experiments are \emph{created}, samples are generated, raw assays collected (e.g. image file, or sequence data), assay processed to get usable data,  experiments \emph{updated} to reflect the new data.
When the data is first generated, it is shared among the team who generated it, so other team members can \emph{read} the data, check its quality, update metadata to indicate data quality, and potentially create additional experiments. Once the research team has validated their initial results, the permissions are changed to make the data accessible to other GUDMAP consortium members, who may further analyze the data and augment the description, while not changing any of the basic experimental data provided by the data generator. For publication, a digital object identifier (DOI) is generated for the data, and may be referenced in a manuscript~\cite{cousijn2018data}. After the journal-imposed embargo period, the access permissions on the data are again adjusted to allow read only access to the public. These changes in permissions happen at different intervals for individual studies, not synchronously across all content in a collaboration.
Hence, from a data and metadata perspective, discovery can be generally structured as repeated and nested cycles of data and metadata creation, reading, and update, i.e. the create, read, update, delete (CRUD)\footnote{While delete may be necessary for pragmatic reasons, in general, all data from a discovery process should be preserved and therefore, we don't include delete in this discussion.} data cycle with the policy associated with CRUD operations changing for each data element over time.

\begin{figure}[ht]
  \centering
  \frame{\includegraphics[width=0.85\linewidth]{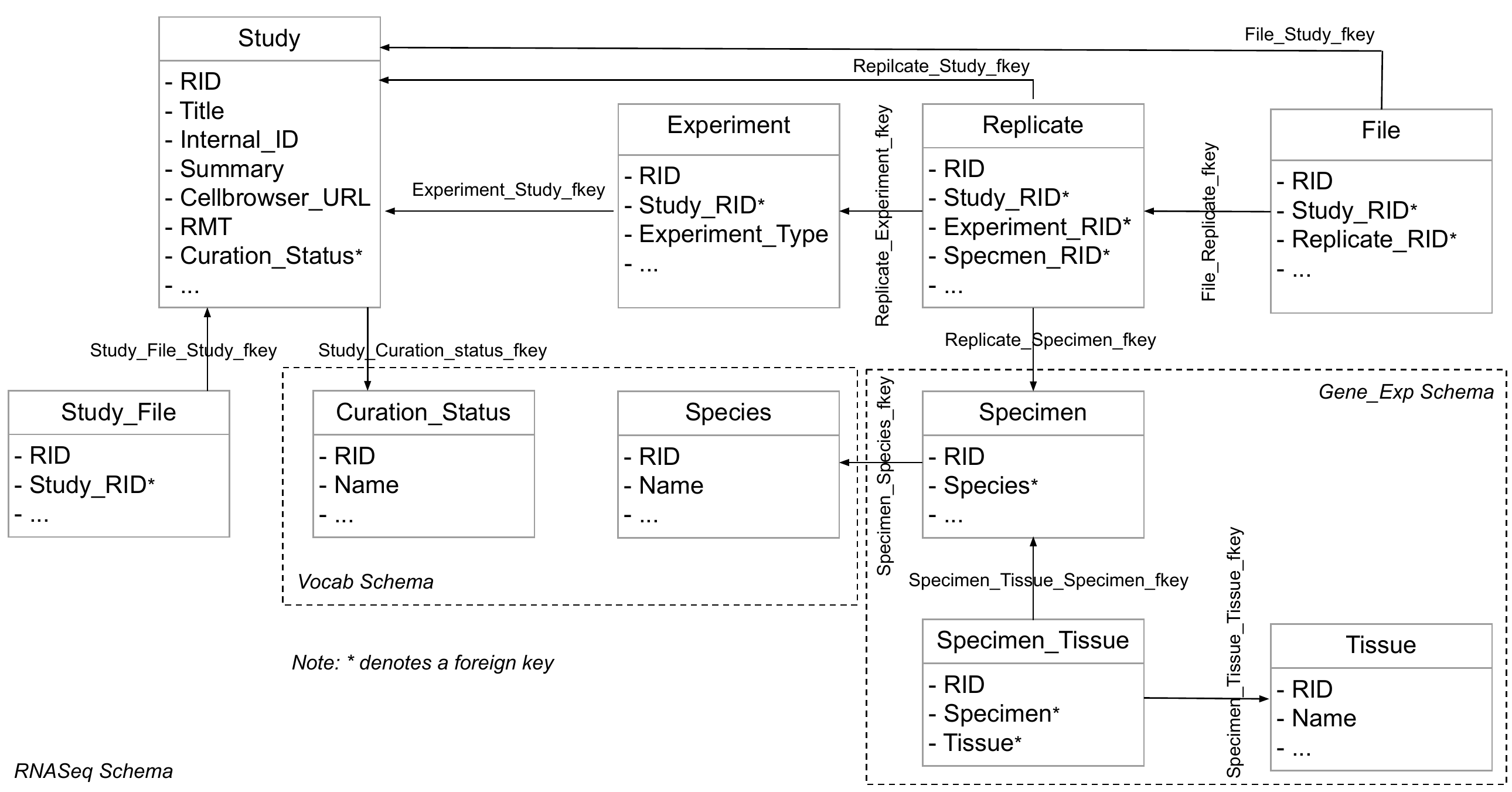}}
  \caption{GUDMAP/RBK data model}
  \label{fig:gudmap-model}
\end{figure}

Figure~\ref{fig:gudmap-model} shows a simplified view of the relational model of GUDMAP/RBK sequencing studies. 
A \emph{Study} such as an RNA Sequencing study composes of one or more {\em Experiment}. 
Each experiment is performed on one or more biological \emph{Replicates} which are used to help validate and quality control the results of the experiment. 
Each replicate is associated with a biosample (\emph{Specimen}) containing biological properties such as {\em Species}, and tissue anatomical sources ({\em Tissue}). 
The metadata of raw and processed files specific to a replicate are stored in a {\em File} table. 
The analysis results done across experiments or replicates are stored in the {\em Study\_File} table.
An important aspect of this model is that there are multiple cardinalities at play (many replicates per experiment, one species per specimen, one replicate per file) which make this data poorly represented by the traditional methods used by scientists to organize data such as spreadsheets or file and directory naming conventions and for which databases are particularly well suited.

While the complexity of this model is necessary to adequately capture the relationship between say replicates, experiments, and samples, it can create challenges from a user interface perspective. For example, even though a user shouldn't be expected to understand all details of the model, it is necessary to create a new replicate and establish the relationship between that replicate and experiment in as simple a way as possible. Another example is that when viewing the details of an experiment, it is likely that the user will want to know what replicates are available and what their characteristics are.


\begin{figure}[ht]
  \centering
  \frame{\includegraphics[width=0.85\linewidth]{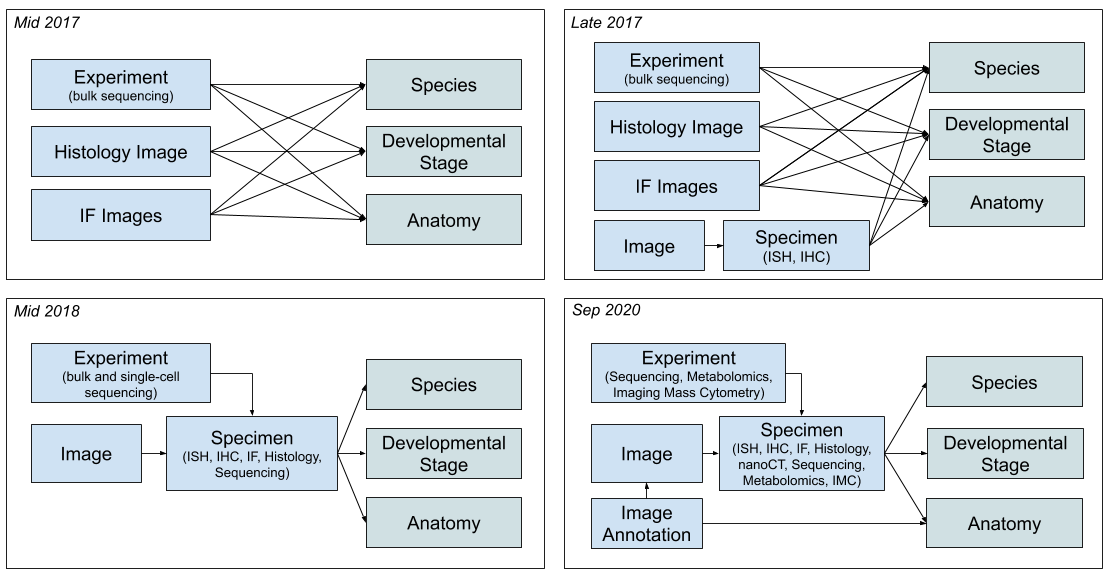}}
  \caption{GUDMAP/RBK model evolution}
  \label{fig:gudmap-model-evolution}
\end{figure}




Another important element of GUDMAP/RBK is the fact that the data model had to change as new data types were required, new technology was developed and use cases were refined. 
For example, the use of single-cell sequencing went from being a promising approach to an essential tool for understanding how gene expression over the period of 12 months~\cite{hedlund2018single}. 
Since the system inception in 2016, the GUDMAP/RBK data model has been \textbf{evolving} to support new data types or use cases.
Figure~\ref{fig:gudmap-model-evolution} shows an example of how a subset of model was evolved. The system was initially designed to support RBK assay types. In mid 2017, the sequencing {\em Experiment}, {\em Histology Image}, and {\em IF Image} schemas independently include metadata that describe biospecimen. In late 2017, the legacy GUDMAP imaging data covering different types of images and a different way of capturing biospecimen was incorporated. In mid 2018, all imaging data are unified and extended to take any type of 2D imaging data from both consortia. We also established a standard model for capturing biospecimen (i.e. via {\em Specimen}), and extended the Experiment model to cover single-cell data. In the latest snapshot (Sep 2020), the Experiment model was further extended to support other types of experiments e.g. metabolomics. The {\em Image Annotation} model has also been added to capture and display annotated shapes within an image.                
     
\subsection{Key Requirements}\label{sec:key-requirements}

The GUDMAP/RBK example illustrates structure and user interaction that is a typical of data-driven scientific discovery. 
We have seen similar patterns in many other scientific use cases~\cite{Bugacov2017} and from these have synthesized what we believe to be key cross cutting requirements for mediating the interaction between users and scientific data repositories in data-driven collaborative efforts.
These requirements were derived iteratively as an integrated part of an agile software development process. Inputs to the requirements synthesis included use case collection and analysis, surveys, and interviews gathered from diverse scientific communities, and continuous production deployments over a 5 year period.

\subsubsection{Provide a collaboratively-maintained metadata repository}\label{sec:req-collaborate}
In the GUDMAP/RBK example, we showed that discovery was achieved by iterative creation, reading, and update of shared data and metadata.
We fundamentally frame scientific data-management as the collection and curation of scientific metadata representing the data byproducts of a project and the scientific context for those byproducts. 
In order to support discovery, these byproducts are tracked and imbued with meaning in a shared metadata repository. In order to collect this metadata and maintain its quality and utility, the participants in the research project who are involved in data production and consumption must collaborate and share responsibility for the structure and content of this repository.

\subsubsection{Provide users with simple consistent interfaces for basic tasks that can be composed to address diverse use cases.}\label{sec:req-web-uis}
The GUDMAP/RBK example shows that diverse data-driven discovery workflows can be decomposed into a small, fixed set of basic subtasks. Perhaps without surprise we have found that create/edit, search and view form a basic set of data-driven operations that can be composed to cover a wide range of models. 
Users should not need deep understanding of the underlying data models nor skill in writing complex search queries in order to perform basic tasks. 
Hence, the details of the interaction interfaces should reflect specifics of the current data model, and the interfaces should compose in such a way as to guide users through the diverse data types available during all stages of scientific workflows and during all phases of projects~\cite{Schuler2014b}. 
This implies that both the details of the subtask and the composition of the subtasks should adapt to the current state of the data model and usage context.

\subsubsection{Support diverse and evolving relational data models and access policies}\label{sec:req-diverse-models}
As illustrated in GUDMAP/RBK, pervasive evolution of data models, data types, and access policy is a distinguishing characteristic of data-driven scientific discovery, as an ongoing process throughout the investigation lifecycle.
The types of data tracked, contextual details, and relationships between data will evolve as a collaboration unfolds. Relational storage models are needed to allow users to share data structuring conventions, represent the intricate connections between related pieces of metadata, and maintain metadata quality while investigation continues. Flexible policy mechanisms are needed to enable different access levels to data depending on a user's role in the system as well as the status of particular pieces of metadata. A collaboration is likely to require evolution of policy to complement their evolving model, but also requires access controls that can track the rapidly changing status of individual experiments and studies at a particular stage of the project's evolution.
Early-phase, exploratory projects need quick setup with simple models and policies while researchers establish experimental protocols, collaboration styles, shared terminology, and data collection standards. Over time, the understanding of a scientific inquiry may mature and along with it, the representation of activities and results must also evolve. Interaction environments for data-driven discovery must adapt to the evolving metadata model and access policies to provide suitable access throughout the lifetime of a data-driven investigation~\cite{madduri2019}.


\subsubsection{Simplify viewing and updating of metadata with complex relationships}\label{sec:req-fk-paths} 
Metadata with many properties and complex relationships, such as species being an aspect of specimens in replicated experiments of studies in GUDMAP/RBK, are a critical aspect of scientific data. 
Tools for creating, editing, and viewing such information in a relational database can be cumbersome if directly exposing all details of the relational model. Interactions with rich models require that applications abstract these properties and relationships and streamline their entry and navigation by bringing together a coherent presentation of relevant data for a specific task such as search and edit without compromising the underlying fidelity of the data model. Useful simplifications of a given relational model require an understanding of the problem domain, and so we require a mechanism for users to enrich their database model with appropriate abstractions and simplifications.

\subsubsection{Support unique patterns of usage and presentation to address evolving community-specific needs}\label{sec:req-annotation} 
In addition to adapting to diverse models, it is crucial to allow research groups to further customize their data presentation. Variant modes of data presentation may be preferred: color-coded quality assessments; statistical summaries; plots, thumbnails, and online previews; download links; or custom visualizations. We observe a need to allow optional, project-specific tailoring of these presentations beyond what the relational model and entity-relationship abstractions can express.


%% file: parts/approach.tex
\begin{figure}[ht]
  \centering
  \includegraphics[width=0.7\linewidth]{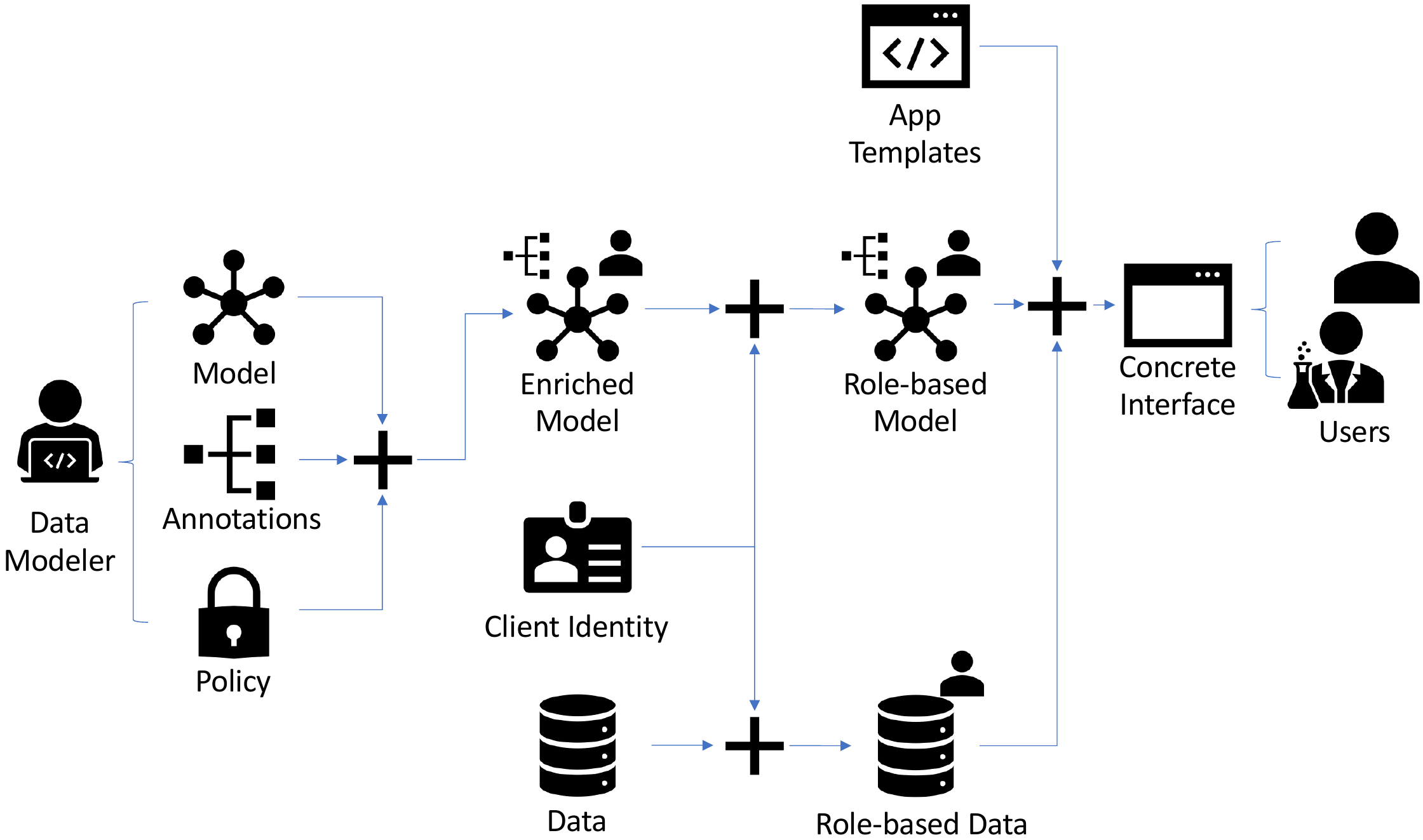}
  \caption{Overview of approach from model definition to concrete interface.}
  \label{fig:approach}
\end{figure}

\section{Model Adaptive User Interface Generation}\label{sec:approach}
In response to the requirements drawn from data-driven science, we have developed a \textit{model-adaptive} approach to interactive applications, in which a small set of interface templates dynamically adapt to the database structure and content to present end users with an interactive experience that accurately reflects the current state of data in an investigation. 
To support our model-evolution and collaborative editing requirements, these template-based interfaces must be generated dynamically throughout the interactive session, adjusting automatically to model and data changes which may occur at any time.

At the outermost level, application templates represent idiomatic interfaces~\cite{cooper2003face} which a user might apply to many parts of their own database, focused on a table, on a table filtered by search criteria, on a specific data record, etc. (These templates are, in practice, modular and decompose into smaller templates oriented towards smaller aspects of database content interaction.) When applied as such, the application-level building blocks map into each of the use case steps we have outlined above, enabling a wide array of complex interactions encountered in data-driven science. The model-adaptive templates translate relationship structures in the database into interface elements, enabling the user to curate, i.e.\ validate and maintain, relationships between data records and to navigate across relationships to incrementally explore or manipulate semantically adjacent content.

Key aspects of our approach are:
\begin{itemize}
    \item Adoption of a relational database platform capable of supporting multi-user access to shared metadata with flexible data access control mechanisms, supporting online model and policy changes (Requirements~\ref{sec:req-collaborate}, \ref{sec:req-diverse-models});
    \item Decomposition of complex workflows into a smaller set of task-specific application templates which can be reused on any database model (Requirements~\ref{sec:req-web-uis}, \ref{sec:req-diverse-models});
    \item Dynamic generation of coordinated and cross-linked user interfaces over the many connected parts of a changing database to provide a coherent data management environment (Requirement~\ref{sec:req-diverse-models}, \ref{sec:req-fk-paths});
    \item Combination of heuristics and user-provided model annotations to abstract the model of a relational database into intuitive entity-relationship concepts which support the basic interaction scenarios (Requirement~\ref{sec:req-fk-paths});
    \item Separation of concerns into a data model, policy, and model annotations, all of which are interpreted and combined to dynamically create an interface (Requirements~\ref{sec:req-collaborate}, \ref{sec:req-diverse-models}, \ref{sec:req-fk-paths}, \ref{sec:req-annotation}).
\end{itemize}

\subsection{Domain-Specific Configuration}
Figure~\ref{fig:approach} illustrates the overall structure of our approach.  A ``Data Modeler,'' possibly a researcher, specifies her domain in the form of a relational \emph{model}, which describes the methods and materials of her research and which will evolve over time. Relational modeling is a common approach to database design; because of our ecosystem of tools and methods focused on model evolution~\cite{10.1145/3400903.3400908} including high-level tools to streamline model evolution in-situ~\cite{10.1145/3335783.3335787}, a modeling approach that emphasizes iterative refinement with increasing levels of detail can be realistically pursued.

In a separate step, model \emph{annotations} may be associated with parts of the model, such as a table or column, to provide additional guidance for interpreting the model. Available annotations can specify a range of simple display directives but also can overlay more sophisticated ER model concepts such as interpreting a chain of foreign key references as an abstracted relationship. While model annotations augment and refer to model elements, they are loosely coupled. They can evolve at a more rapid rate than the underlying data model, and they may be maintained by a different user, who is more concerned with use case specific interaction issues.

The final configuration step is to classify the users of the system in terms of role-based access control {\em policies}. The policies may be used to prevent some users from viewing or editing particular entity types, particular entities, particular attributes of an entity type, or particular relationships. A common need in scientific use cases is to be able to have access control which varies over time and which is data dependent~\cite{8109188}. For example in initial phases of an investigation, data may be restricted to a small group of investigators, later shared across a research consortium, and eventually made available to a broad research community once pre-publication data embargoes are lifted. This requirement dictates that the generated interfaces adapt to both the user's role and the data access policies.

\subsection{Generic Entity-Relationship Interactions}
Our interaction templates are model-driven, meaning that they interpret database structure and adapt, not encoding any fixed assumption as to what model should exist. These templates are designed for specific modes of presentation or interaction which are suitable for our data-management use cases, generalized as manipulation of entity-relationship~(ER) structures we can derive from the database model.
We have designed interaction elements for filtering sets of entities by search criteria, listing sets of entities, viewing details of an entity, creating and editing specific entities, creating and destroying relationships between entities, viewing sets of entities related to a single entity, choosing entities from a listing, etc.

Because we provide interaction over a relational database rather than a more abstract entity-relationship store, we must bridge the gap between the extant relational database model and the desired ER concepts. This is not a fully formalized ER model, but rather a set of ER vignettes providing just the aspects of the ER model needed by the interaction templates. We use heuristics to map the database model onto these concepts, recognizing certain idioms for how relationships are encoded in relational databases. For our purposes, relational tables simultaneously correspond as entity types and as entity sets, and scalar properties of an entity type can be derived from the columns of a table. Thus, creation or editing of entities corresponds to creation or editing of rows in a table. Search of an entity set is search over rows in a table. Relationships correspond to foreign key references or chains of such references. Our interaction templates consider an abstracted list of relationships, abstracting over the foreign key constraints involving the table.
To provide further options for customization and guidance in these mappings, we use model annotations. These allow users to provide further hints to enhance or override our heuristic interpretation of their database model.

When instantiated, each generic template produces an interface element. Many of these templates are model-driven composites, translating the model structure into isomorphic interface structure, where nested templates are used to translate individual parts of the model. For example, the properties of an entity type inform the input fields in a data-entry form. Others are data-driven composites, translating query results into display values. For example, a variable-length query result instantiates a variable-length result listing. Of course, many templates are simultaneously model-driven and data-driven, as mapping of structured data into an interface requires an awareness of the data structure and content. For example, a set of entities in a tabular listing would use a table structure derived from the shared entity type, iterating over the number of entities in the result set, and appropriately mapping the datum in each field of each entity into a cell in the tabular representation.

\subsection{Interface Customization}
Model annotations are used to customize the interfaces generated from the database model. Here, we provide a brief summary of concepts necessary to understand our approach. In subsequent sections, we will expand on these topics in more detail.

Simple display directives may introduce small amounts of content denormalization or reformatting, such as formatting of numbers or dates, converting raw URLs into an inline thumbnail display versus a download link, or splicing together several fields based on an interpolation pattern.
The list of properties displayed for an entity type can be reordered or selectively suppressed via an annotation. Similarly, the list of displayed entity relationships can be customized to reorder or suppress relationships. For complex cases, a chain of foreign key references can be abstracted as a named relationship and supplied to augment the list of available relationships understood by the interface.
These can also be used to augment the list of properties, i.e.\ a relationship to another entity can be presented as a logical property of the entity, or a set of related entities presented as a nested set. Similarly, aggregate calculations over related entities can augment the entity with a computed result.


An important variant of the property list for an entity type is a list of search controls or \emph{search facets}. While heuristics can infer a default list of facets by combining the list of properties and list of relationships, an annotation may also customize this list to reorder, suppress, or augment the choices. In this way the user can enable and promote useful search controls over the most relevant aspects of an entity.

\subsection{Interface Synthesis}
Together, the three elements of domain-specific configuration (schema, annotations, and policy) create an {\em enriched model}. Combining this with information about user identity and role, our framework converts the enriched model into a {\em role-based model}---a model decorated with the user's data-access privileges (create, read, update, delete) and pruned of any columns, tables, or foreign key references the user is not allowed to view. With this, the interaction templates can selectively disable or hide structures and controls depending on the user role.

The storage platform provides access to database content that has been structured and filtered to produce {\em role-based data} conforming to the role-based model. Together, these byproducts are supplied to the interaction templates to generate a fully concrete interface providing a given user with appropriate access to specific database content.
This dynamic combination of configuration, policy, user identity, and interaction templates provides an interface that constantly adapts to changes in various deployments. This approach allows the reuse of the technology to support multiple user communities without repeated software development to adapt services and applications for each project and model revision.

\subsection{Data denormalization}\label{sec:denormalization}
Modeling techniques that can improve accuracy and integrity of data often complicate issues of search and display, in that information best presented together may be separated by one or more relationships. While it is considered a best practice to design a well normalized model, interacting with highly normalized data is difficult and can degrade the user experience. Often, the solution to this is to resort to relaxing the data normalization practice, which sacrifices data fidelity or increases model complexity. Our approach provides the opportunity to denormalize data on the fly to improve the user experience without sacrificing model accuracy. 

In the simplest case, our interaction templates understand zero-or-one and exactly-one cardinality relationships encoded as foreign key references in the relational model. We represent the single related entity as a logical property named by the relationship and present the name of the related entity as the property's value. Editing features can then provide an entity-selection interaction to choose appropriate values. The entity name is denormalized content retrieved from the referenced entity. Heuristics can find typical entity properties which serve as entity name, e.g.\ fields like {\em title}, {\em name}, or {\em accession\_number}, but this can also be customized using model annotations. This form of denormalization serves two common user intents: representation of properties governed by {\em controlled vocabularies} and designation of a {\em parent} entity with respect to a specific {\em child} entity. These are two alternate ways of thinking about the identical data structures, differing only in the intent behind the referenced entity set (for example, the experiments in a study and the specimens of a species use the same simple foreign-key structure in the database model in Figure~\ref{fig:gudmap-model}).

Two closely related cases are zero-or-more cardinality relationships encoded idiomatically as a single foreign key reference or as a binary-association table. We can represent the set of related entities in several forms, including lists of entity names and nested tabular listings. These relationships can be rendered as complex properties of an entity or in summaries of related entities for a given, {\em focal} entity. Editing features vary by the relational encoding. The foreign key scenario is a set of {\em child} entities for a given {\em parent} entity, an inversion of the simple case described above, and we can provide interactions to create new child records or direct the user to edit an existing child record to change its parentage. A binary association table captures many-to-many peer relationships and we can provide interactions starting from an entity on either side of the relationship to prune connections or to make bulk revisions using an entity-multi-selection interaction.

As a generalization, we can also present user-defined relationships which abstract over other chains of foreign key references in the database. For read-only interactions, these can be presented like the previously described relationships. However, we cannot support concise relationship editing since there would be ambiguity as to how to change the database state. Editing of such structured content requires a less abstract ER interpretation, more closely aligned to the actual relational storage model. For example, the individual links in the chain of foreign key references are each embedded in tables, and these tables can be interpreted in simpler parent-child entity vignettes to incrementally create or revise intermediate records in the chain. In fact, if instructed to do so, our application templates would also represent a binary association table in such a way---it is simply an entity with two entity-typed properties, relating it to the entity on either side of the binary association.

All of the preceding relationship types can also be used to denormalize related entities as search facets. Search features can provide an entity-multi-selection interaction to set search criteria with respect to each named relationship. As a variation for both search and read-only presentation, we can also denormalize other related entity content such as scalar properties, rather than representing a set of entities in the user interface. Search features can provide appropriate interactions to control search criteria based on these related scalar values, e.g.\ range selectors for numbers and dates. For presentation, some interaction templates can directly represent sets of related scalar values. As a final option, we can expose computed aggregates over sets of related entities or related scalar values, representing the computed result of a chosen function (e.g. count, minimum, maximum, sum) as another scalar property. Earlier, in Section~\ref{sec:use-cases}, we illustrated some of these denormalization concepts with respect to an actual problem domain.

%% file: parts/applications.tex
\section{Application Templates}\label{sec:apps}

Rather than support arbitrary task models~\cite{Da-Silva2000a,Paterno2001a}, our approach is to decompose user interactions with data into a small suite of interrelated application templates which are connected by relationships found in the structured data and composed to support a broad range of user scenarios. Based on our use case analysis, we have chosen to define three interrelated application templates focusing on specific tasks of data-entry, searching, and viewing, decomposed into additional modular templates which provide reusable components and consistent interaction styles throughout. These templates can be instantiated with respect to any table in the database, though in many projects the user will start with templates focused on a shorter list of important entity types and only visit the subset of tables reached by navigating the important relationships exposed by the dynamic interface generator.

\begin{figure}[htb]
\begin{subfigure}{.49\textwidth}
  \centering
  \includegraphics[width=\linewidth]{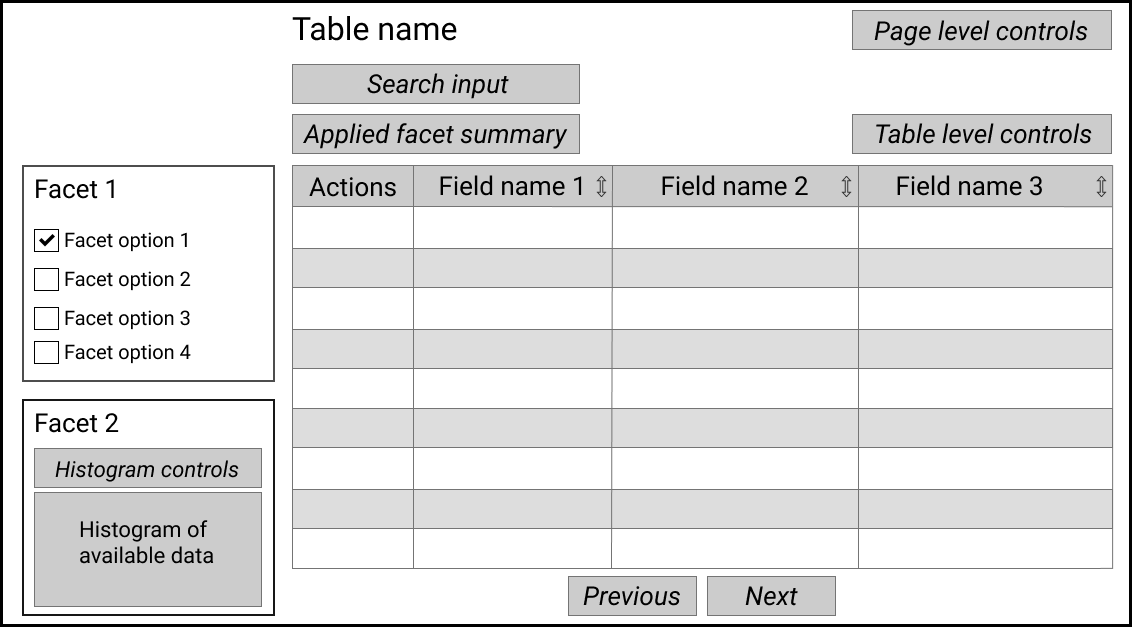}
  \caption{Faceted search template with filter controls and a result set drawn from a focal table, i.e.\ entity set.}
  \label{fig:recordset-abstract}
\end{subfigure}
\begin{subfigure}{.49\textwidth}
  \centering
  \includegraphics[width=\linewidth]{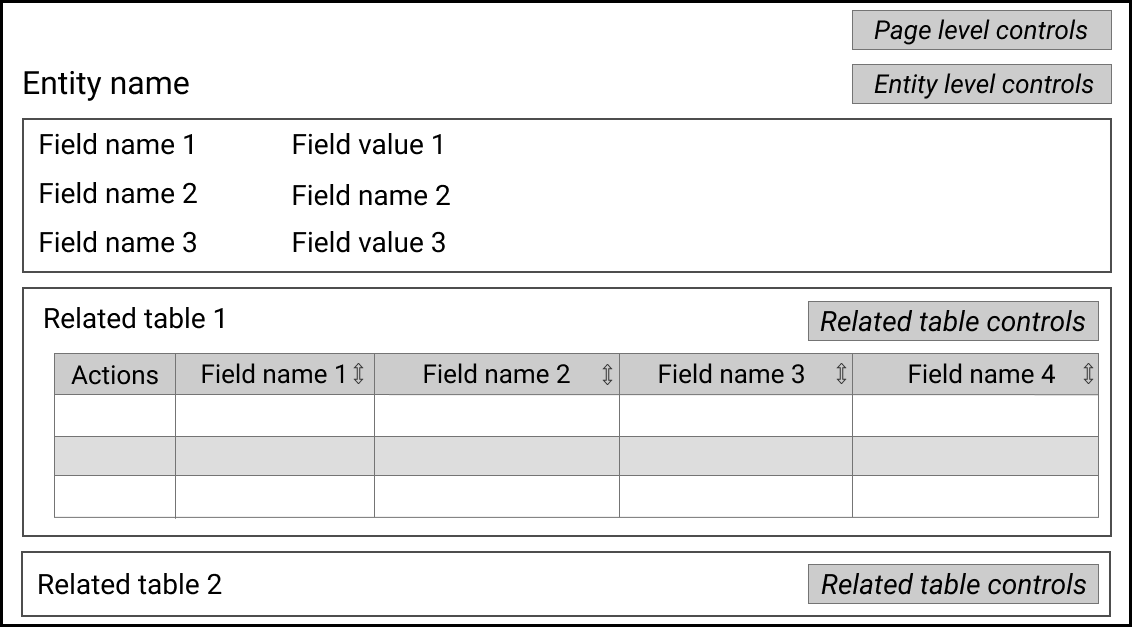}
  \caption{Detailed template for a focal row, i.e.\ entity, with summaries of related entities.}
  \label{fig:record-abstract}
\end{subfigure}
\begin{subfigure}{.49\textwidth}
  \centering
  \includegraphics[width=\linewidth]{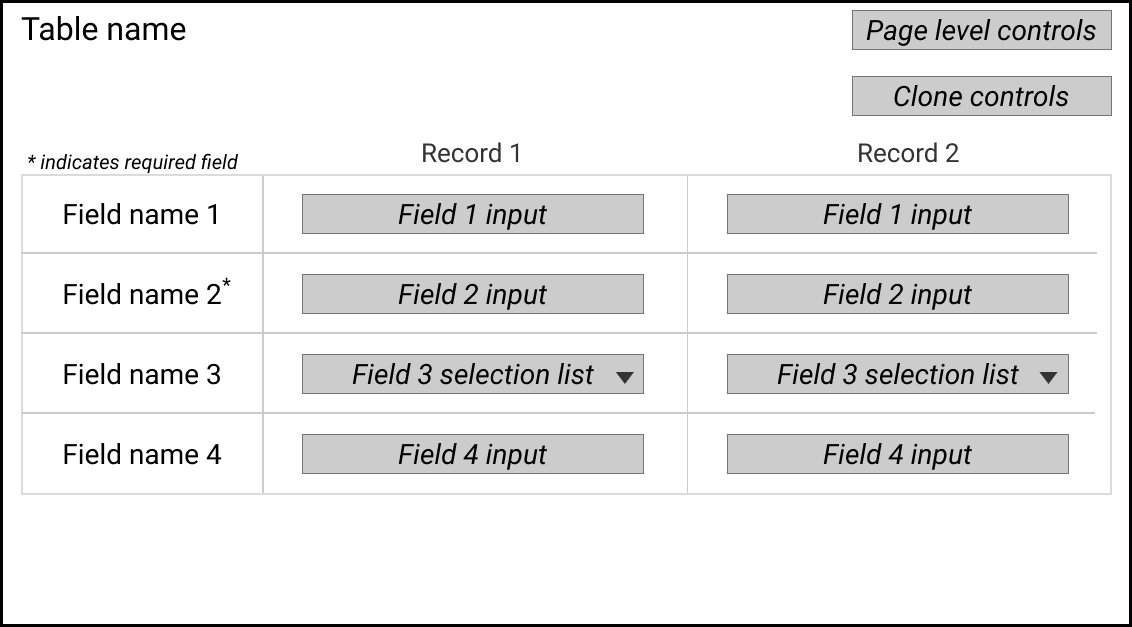}
  \caption{Data-entry template for a focal table, i.e.\ entity type. \\}
  \label{fig:recordedit-abstract}
\end{subfigure}
\begin{subfigure}{.49\textwidth}
  \centering
  \includegraphics[width=\linewidth]{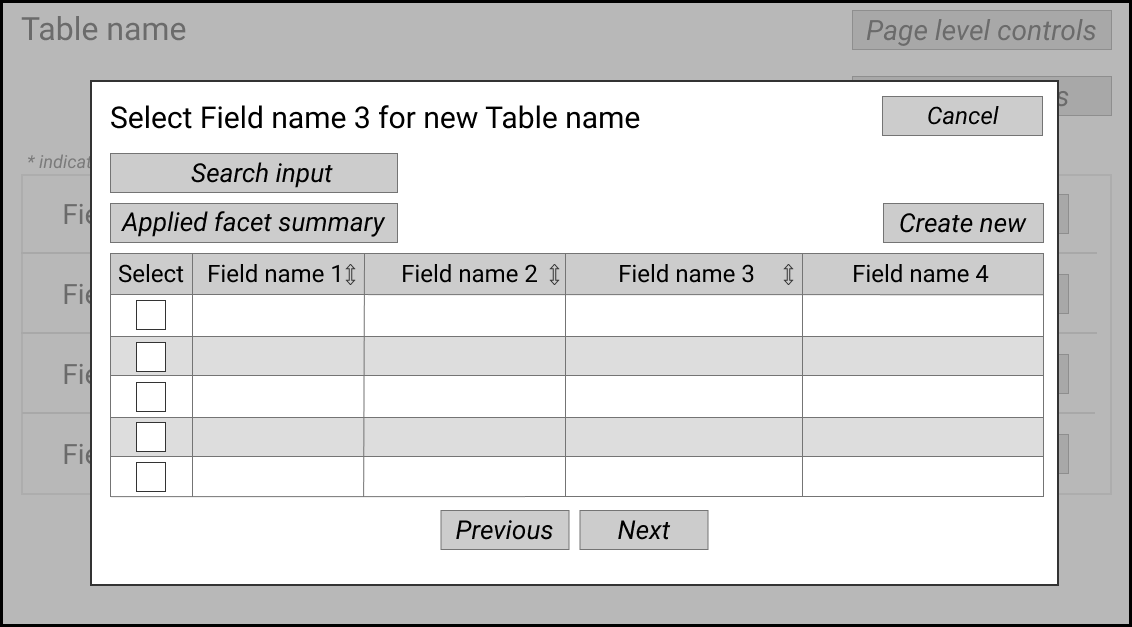}
  \caption{Row, i.e.\ entity, selection template with choices suitable for a specific foreign key reference, i.e.\ directed relationship.}
  \label{fig:recordedit-fk-abstract}
\end{subfigure}
\caption{Application templates for search, viewing, data-entry, and selection while applying an entity-relationship interpretation to a relational database.}
\label{fig:chaise-abstract}
\end{figure}

These three templates provide the fundamental CRUD operations needed to support a collaborative ``mini-workflow'' of "search, view detail, and if not found, create, or if incorrect, edit or delete" associated with individual entities. Supporting interactions allow the relationship structures of the database to be managed and exploited.
The search template focuses on one entity type and shows a set of entities matching search criteria, offering controls to switch into per-result templates. The detailed entity view template focuses on one entity and provides summary information about sets of related entities linked by relationships. The template offers controls to switch into a data-entry template on the same focal entity, to a drill-down search of a set of related entities, to a detailed view of one related entity, etc. The data-entry template also supports starting with blank forms for a given entity type in order to create new content from scratch. A common modular function is an entity-selection template, shared by editing and search interactions.

Figure~\ref{fig:chaise-abstract} illustrates the abstract templates for these tasks. These application templates are configured and delivered by the method described in Section~\ref{sec:approach}.  
Each application template defines a page structure when instantiated for a focal point of the model and any required data focus, e.g. the detailed entity template in Figure~\ref{fig:record-abstract} focuses on one row of one table and follows that with summaries of content related to that row.

\subsection{Presentation Contexts}\label{sec:context}


Across our application templates, we have identified a set of significant {\em presentation contexts} where different interaction modes may benefit from different interface customization directives. To support such selective reconfiguration, we define a set of context names, including:
\begin{itemize}
    \item {\em detailed}\/: A focused view of one entity.
    \item {\em compact}\/: A listing of multiple entities, such as a search result.
    \item {\em row\_name}\/: A concise reference to one entity by human-readable name.
    \item {\em entry}\/: A data-entry form for one entity type.
    \item {\em filter}\/: A collection of filter controls for searching an entity set.
    \item {\em *}\/: A default configuration for many contexts.
\end{itemize}
For convenience, we define a ``*'' default context which can be used to share a common configuration across multiple presentation contexts, for those situations where differentiated configuration is not felt necessary. Conversely, we also define addition sub-contexts (not listed) to allow narrowly-targeted reconfiguration for certain modes, e.g.\ a data-entry form might benefit from differentiated configurations when creating entities versus when editing existing entities. Configuration search rules allow a user to mix and match some of these context-specific configurations without being forced to overly specify instructions where the defaults are satisfactory.

Many of these presentation contexts are closely aligned with the major mode of an application template. However, a particular application template can contain several modular templates, each of which exhibits a distinct presentation context or in fact might activate/deactivate such auxiliary presentation contexts as user-driven interaction is performed.
The model-driven interface generator can determine an entity-relationship interpretation of any table in the model for any desired presentation context, as required to process and instantiate interaction templates. This allows for very precise and selective configuration of interfaces by annotating specific parts of the model with contextualized instructions.

\subsection{Data-Entry Template}

The data-entry template~(Figure~\ref{fig:recordedit-abstract}) enables users to {\em create}, {\em update}, or {\em delete} an entity or set of entities in the database. Data-entry forms are generated based on the interpretation of the storage table as an entity type, with appropriate input fields for inferred properties of that entity type. It uses the {\em entry/edit} or {\em entry/create} contexts to determine the configuration when producing this property list.
Likewise, {\tt NOT NULL} constraints are translated into {\em required} properties, and suitable data-entry tools are offered for typed scalar properties such as numbers, dates, and timestamps. Relationships encoded as foreign key references in the table are mapped as properties containing a selected entity. Such controlled inputs supplant free-text entry and prevent many accidental data errors.
The template can also pre-validate entries to alert users to likely errors, or surface errors detected by the relational database storage platform.

Available and selected foreign entities are displayed with human-readable entity names while populating actual foreign key fields with their often cryptic values. 
During entity-selection, users can follow an auxiliary flow, if desired, to extend the set of available entities. This can be used to add a new term to a controlled vocabulary and immediately choose the new term in the data-entry template.
For further customization, the entity-selection template can be configured to filter choices based on the earlier user inputs in the form, smoothing some data-entry tasks with many interrelated choices.


We provide a bulk variant of the data-entry template. When creating entities, a user can incrementally expand the form to edit multiple entity descriptions at once and request that they all be created together. When accessed by cross-links from other application templates, bulk-editing of multiple existing entities can also be performed. The forms also adapt to access rights, e.g.\ disabling input on fields immutable for a given user or completely hiding fields that a user is not allowed to see. This adaptive data-entry form for structured data addresses drawbacks of typical approaches to managing scientific data---like shared spreadsheets, where collected data often accumulates with latent errors and omissions.

\subsection{Faceted Search Template}

The faceted search template~(Figure~\ref{fig:recordset-abstract}) provides structured filter controls and basic text search over a table using an entity-relationship interpretation. The facet panel contains a list of filter controls, providing appropriate filter predicates for either basic entity properties, basic properties of entities linked by a given relationship, or identity of linked entities by a given relationship. It uses the {\em filter} context to determine the configuration for this panel. As users select or deselect the predicate values of a facet, search results and potential values of other facets are automatically updated. Search results are displayed as tabular data. It uses the {\em compact} context to determine the configuration of the search results table. The results template shows a subset of entity details and provides per-result navigation to detailed entity views or data-entry views. Certain bulk operations are also offered, such as deletion of entities found by a particular search criteria, or a bulk-editing variation of data-entry.


\subsection{Entity Viewing Template}

The entity viewing template~(Figure~\ref{fig:record-abstract}) presents details of a focal row of one table using an entity-relationship interpretation which also includes summaries of {\em  related entities}---entities linked to the focal entity by named relationships. It uses the {\em row\_name} context to determine the configuration for the focal entity's name, the {\em detailed} context to determine the configuration of the focal entity's property list, and the {\em compact/brief} context to determine the configuration of each set of related entities. An auxiliary interaction allows management of certain relationships as described in Section~\ref{sec:denormalization}.

The related entity sets are usually displayed as embedded tabular sets, but this can be customized through model annotations.
Scientific data visualization, such as plots or 3D rendering, can be flexibly integrated into the display via a customized markup rendering engine and hypertext \texttt{iframe} elements embedded in the display. (An illustration of this appears under "Cell Browser" in Figure~\ref{fig:gudmap-study-record} in our subsequent implementation description in Section~\ref{sec:implementation}.) 


\subsection{Workflow Composition}
Often, users start with search or data-entry templates for a small number of important entity types. Then, more complex workflows branch off as users traverse navigable links or activate editing tools embedded in subsequent views. This approach allows a collaboration to synthesize new workflows from these reusable tools. As new projects are formed or an existing project evolves, we avoid the repeated engagement with UI developers that is typical of conventional approaches. 

\begin{figure}[ht]
  \centering
  \frame{\includegraphics[width=1.0\linewidth]{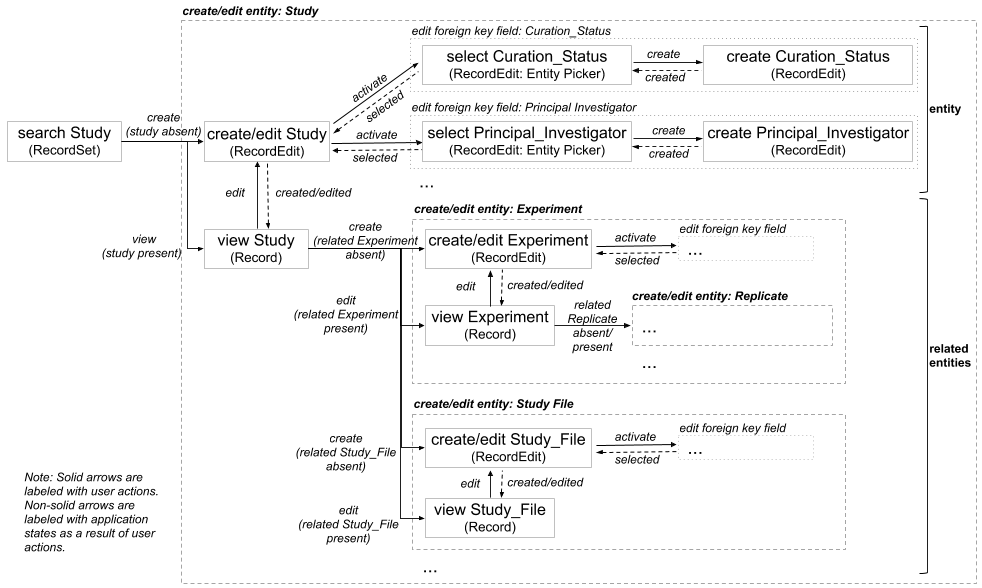}}
  \caption{Curation workflow of GUDMAP/RBK sequencing data}
  \label{fig:gudmap-workflow}
\end{figure}

We demonstrate the role of these three application templates in composing a complex curation workflow through our GUDMAP/RBK example (Section~\ref{sec:use-cases}). Figure~\ref{fig:gudmap-workflow} and ~\ref{fig:gudmap-model} show a typical sequencing data curation workflow and the relevant model. A sequencing {\em Study} submission is completed when its core entity and related entities are properly created. The same rule is (recursively) applied to each related entity. These basic steps form building blocks for composing complex workflow involving arbitrary number of tables.  As a collaborative environment, the database may reflect incomplete work depending on the integrity constraints chosen by the data modeler.

A typical workflow starts with a lab member searching for their on-going study in {\em Study} search (illustrated in Figure~\ref{fig:gudmap-study-recordset} in Section~\ref{sec:implementation}). If their study isn't found, they then click the "Create" button on the search page which will direct them to a data-entry page with a blank Study form to create a new entity (Figure~\ref{fig:gudmap-study-recordedit} in Section~\ref{sec:implementation}). While entering a foreign key field (e.g. Curation\_Status), they search for a specific entry. If the entry is not found, they can then branch off to create a new entity in the foreign key table by clicking "Create New". This action will open another data-entry tab which leads to its own entity creation workflow. In many cases, the foreign key entities are representing controlled vocabulary terms that become terminal leaves in this recursive workflow.

In Figure~\ref{fig:gudmap-workflow}, we depict the common foreign key editing process with "edit foreign key field" boxes. After the foreign key entity is created, the user can select the newly created entry, then proceed to complete the rest of the Study form. After a Study entity is created, a detailed entity view with empty related entity sections is shown (since the user has not yet populated these relationships for the new Study). The user can then create related entities (e.g. Experiment and Study\_File) by clicking "Add" for the corresponding relationships. The "Add" button will direct the user to another data-entry tab for creating the corresponding related entity (e.g. Experiment) with the parent Study already filled in. The same curation workflow---creation of an entity and its related entities, depicted as "create/edit entity" boxes---is repeated for all related entities which form part of this new Study. An alternative flow to creating/editing a related entity is to initiate a search in the appropriate entity type by desired criteria, and then activate an edit view on the found result. This is useful in curation, e.g.\ when an Experiment has been erroneously assigned to the wrong Study, but can be found by other characteristics and edited to relate it to the correct parent Study.

To revisit a partially completed Study, the user can first search for their study, then navigate to its detailed entity view. Once there, they can either edit the existing Study entity, add new related entities, or edit/delete existing related entities. The user can traverse a sequence of related entities to navigate to a specific part of the workflow (e.g. Study->Experiment->Replicate) to fill in missing details.   
The preceding example is a typical top-down approach. Different variations of the curation workflow can also be applied to achieve the same goal e.g.\ creating all Specimen entities in bulk first, then proceed to create a study, an experiment, and all related replicates in bulk.   
A Study entity is completed as long as its corresponding Experiment and Study\_File entities are completed, which implies all Replicate entities associated individual experiments, and all File entities associated with individual replicates are completed.         


%% file: parts/annotations.tex
\section{Annotations}\label{sec:annotation}
\label{sec:annotations}

Annotations are a core element of the enriched model. Consumed by an application template, they are used to configure the application to the use case at hand. Annotations guide the entity-relationship interpretation of the underlying relational database. As discussed above, in the absence of a user specified annotation, an application template will consume a heuristically-generated, default ER interpretation.

An annotation is associated with a specific element of the relational model and hence is scoped to a database, schema, table, column, key, or foreign key. 
Annotations are captured in an extensible key-value store associated with the database model, hence a model element, such as a column, may have one or more annotations that consist of the annotation name, and a value whose meaning will depend on the type of annotation.  Some annotations are further structured by presentation context names~(see Section~\ref{sec:context}) to allow multiple, differentiated configurations for different sub-components of the user interface.
Because an annotation is a mapping over the relational model, the writer of an annotation must have some knowledge of at least elements of the relational model, and the annotations must evolve in conjunction with the underlying model. 
To aid in this, we have developed tools for validating consistency between annotations and the underlying model as well as creating tools that co-evolve a model and corresponding annotations.

In practice, we have found that separation of modeling from presentation and using a simple key-value representation for annotations to be beneficial both from the perspective of flexibility and compartmentalization of skill sets (data modeling, vs. presentation).
We typically will see many cycles of evolution of annotations to adapt the interface to changing use cases and end-user feedback, leading to further refinement of customized settings.
The incremental refinement of presentation may itself be driven by the evolving needs of the discovery process, and hence a sequence of annotation modifications may lead to the need for evolution of the underlying database model, causing the entire process to repeat.



While the set of potential annotations is open, in practice, we have found that a small number of annotations that fit into four basic classes: table specific, column specific, foreign key specific, and generic annotations have proven effective in covering many different use cases. 
Table~\ref{tab:annotations} summarizes annotation keys representing different UI customization themes. This annotation set is not definitive, rather what we have found to be useful for addressing arising use cases. In Section~\ref{sec:deployments}, we provide detailed analysis of how annotations are used in practice.

\begin{figure}[ht]
\begin{subfigure}{\textwidth}
  \centering
  \frame{\includegraphics[width=\linewidth]{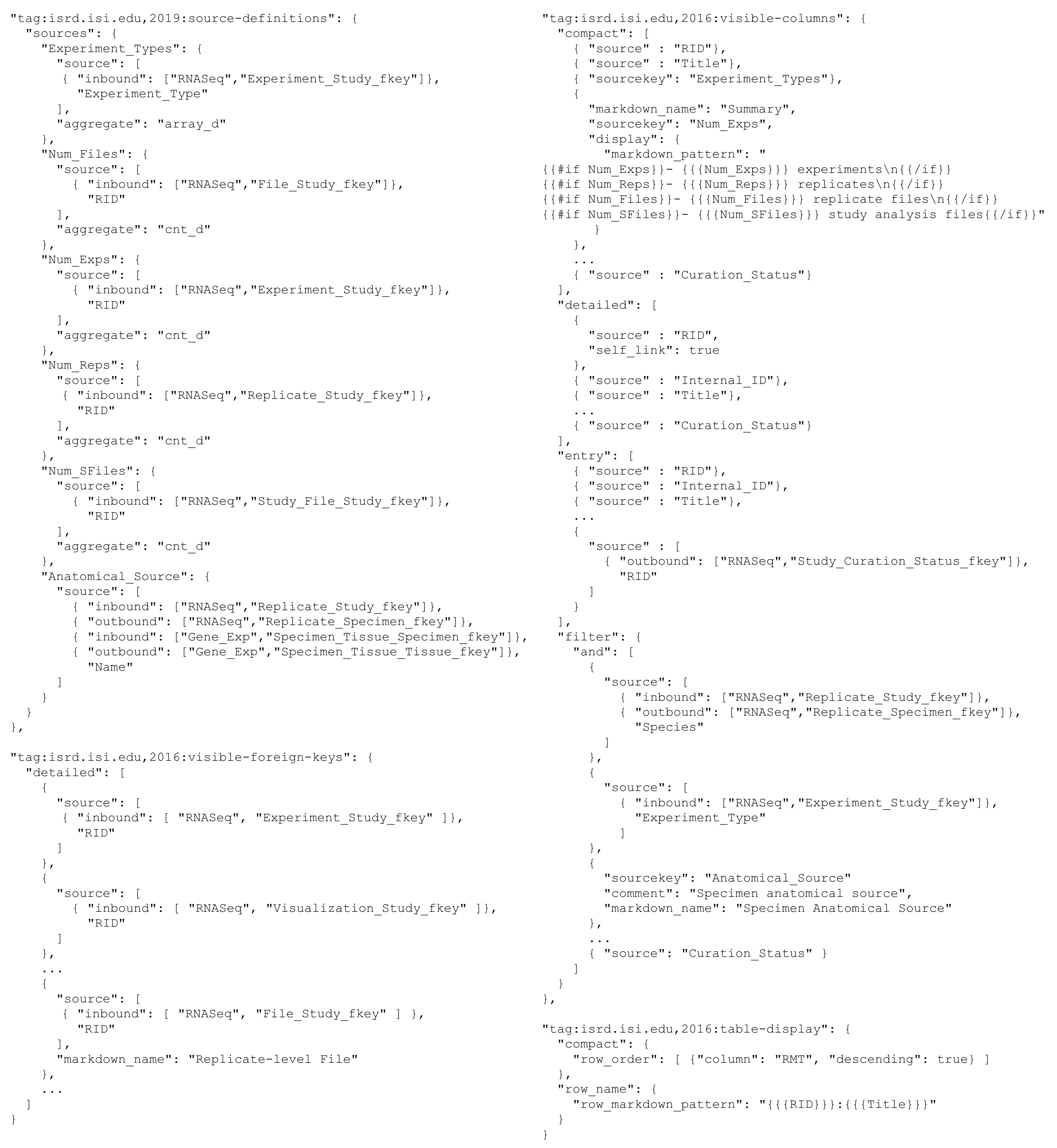}}
  \caption{Table annotations}
  \label{fig:table-annotations}
\end{subfigure}
\begin{subfigure}{\textwidth}
  \centering
  \frame{\includegraphics[width=\linewidth]{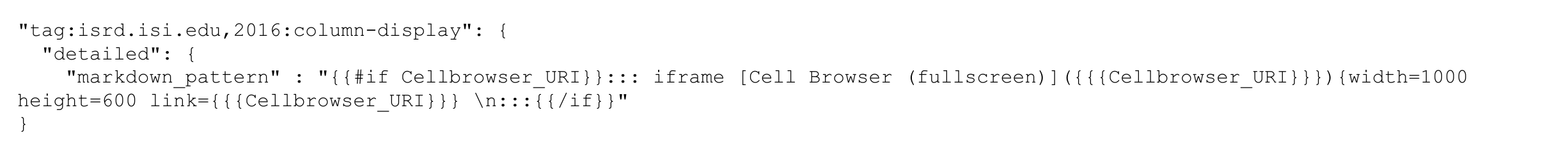}}
  \caption{Column annotations}
  \label{fig:column-annotations}
\end{subfigure}
\caption{GUDMAP/RBK Study related annotations}
\label{fig:annotation}
\end{figure}

Figure~\ref{fig:annotation} shows examples of table and column annotations associated with the GUDMAP Study table (see model in Figure~\ref{fig:gudmap-model}). Figures ~\ref{fig:gudmap-study-recordedit}, ~\ref{fig:gudmap-study-recordset}, and~\ref{fig:gudmap-study-record} capture the actual displays of the Study model as configured by these annotations and rendered by our implementation, discussed later in Section~\ref{sec:implementation}.

\begin{table}[ht]
  \caption{Annotation Summary}
  \label{tab:annotations}
  \begin{tabular}{p{0.26\linewidth}p{0.7\linewidth}}
    \toprule
    Annotation Keys &  Summary \\
    \midrule
    {\em table specific}: \\
    source-definitions & Consolidated property and relationship definitions. \\
    visible-columns & Mapping of columns as entity property list. \\
    visible-foreign-keys & Mapping of referential structure as relationship list. \\
    table-display & Entity naming, sorting, and assorted display options. \\
    \midrule
    {\em column specific:} \\
    column-display & Display options for column content. \\
    asset & Mapping of columns for a file-oriented property. \\
    required & Emulate a not-null constraint for interactive users. \\
    \midrule
    {\em foreign key specific:} \\
    foreign-key & Display of foreign key names and selection constraints e.g. show only purification protocols for protocol selection. \\
    \midrule
    {\em generic:} \\
    display & Relabel of names and comments (tooltips) of model elements.  \\
    generated & The associated model elements are generated; disable user input.  \\
    immutable & Once set, the model elements cannot be changed; disable editing. \\
    \bottomrule
  \end{tabular}
\end{table}  

\subsection{Table-Specific Annotations}

Table-specific annotations are only meaningful when applied to a table in the model. They influence the overall interpretation or presentation of the table as an entity type.

\subsubsection{Source-Definitions} The {\tt source-definitions} annotation allows a consolidated set of definitions of entity properties and relationships, so that they can be reused (by reference) in the {\tt visible-columns} and {\tt visible-foreign-keys} annotations. This allows the modeler to avoid repeated inline definitions when they want to reuse the same concept in multiple, contextualized lists. Each definition can involve traversal of chains of foreign key references (optional) and mix-in other choices such as whether to provide an entity interpretation of related records, a computed aggregate, or projection of scalar values as discussed in Section~\ref{sec:denormalization}. Some of these options may only be appropriate for a subset of all possible presentation contexts. A property defined here or inline in a property list may include custom display instructions and can draw from more than one designated property or relationship, e.g.\ presenting a "minimum to maximum" range as a property instead of two separate properties for minimum and maximum values.

\subsubsection{Visible-Columns} The {\tt visible-columns} annotation configures the mapping of the table definition into an ordered list of entity properties for each presentation context. (It gets its name from the heuristics which derive a default property list from the columns of the table definition.) Each list can contain inline definitions of novel properties but also may reference definitions in the {\tt source-definitions} annotation. In practice, this annotation affects the structure of facet control panels, tabular listings, and detailed entity views.

\subsubsection{Visible-Foreign-Keys} The {\tt visible-foreign-keys} annotation configures the mapping of the table and adjacent model definition into an ordered list of relationship types between this entity type and others. (It gets its name from the heuristics which derive a default relationship list from the set of foreign key constraints involving the table.) Each list can contain inline definitions of novel relationships but also may reference definitions in the {\tt source-definitions} annotation. In practice, this annotation affects the structure of related entities summaries and also may influence facet control panels when they are not fully configured by an explicit visible-columns list.

\subsubsection{Table-Display} The {\tt table-display} annotation configures entity-naming for rows of the table, default sort criteria for ordered listings of multiple entities, and assorted other display options. An alternative display can be configured to replace the default tabular view with a different presentation such as a bullet list, comma separated list, or even a custom display such as a heatmap visualization. 

\subsection{Column-Specific Annotations}

Column-specific annotations are only meaningful when applied to a column in the model. They influence the interpretation or presentation of the column as a property of an entity type.

\subsubsection{Column-Display} The {\tt column-display} annotation configures custom presentation rules for the content of a column when used as a property in different presentation contexts. These influence property definitions which project the column (unless those property definitions override with their own custom display options). Customization examples are displaying multiple properties together (e.g. first name and last name); formatted text (e.g. display yes/no for Boolean values); linkable URL or inline image display; styled text (as demonstrated by Curation Status column shown in Figure~\ref{fig:gudmap-study-recordset} and \ref{fig:gudmap-study-record-details}); or custom visualization (as demonstrated by Cell Browser on Figure~\ref{fig:gudmap-study-record-details}). By default, the raw values are formatted and displayed based on their data types e.g. thousand separated integers, {\tt YYYY-MM-DD} formatted date. 

\subsubsection{Asset} The {\tt asset} annotation configures a mapping to interpret columns related to digital asset management, i.e.\ file management, as an integrated set of asset interaction templates. The mapped columns identify the stored URL as well as other properties of a file (file name, file size, checksum). Our interaction templates can provide a download link and expose other information about the file, or even enable interactive file-upload for use cases where users submit files individually while curating metadata.

\subsubsection{Required} The {\tt required} annotation configures an emulated not-null constraint on a column. This has proven useful where a data modeler has had to permit null values in a database to accommodate legacy data or special cases coming from automated processing pipelines, but wants to encourage new data submissions by interactive users to always include certain properties. Rather than being enforced in the database platform, the enforcement occurs in the validation logic of the interaction templates.

\subsection{Foreign-Key-Specific Annotations}

One foreign-key-specific annotation is understood by our current templates, influencing our heuristic interpretation of foreign keys as relationships.

\subsubsection{Foreign-Key} The {\tt foreign-key} annotation configures the relationship name for relationships inferred from underlying foreign key structures in the database. It supports directional naming to name the entity or set of entities that would be considered related entities to an entity on the opposite side of the relationship.  
Another usage of this annotation is to specify the filtering criteria of the selection list displayed in an entity-selection interaction. For example, while editing an experiment entity, under a Purification Protocol property, only protocols under that category should be shown.   

\subsection{Generic Annotations}

Generic annotations are applicable at more than one level of the model, i.e.\ they can customize behaviors of entities when applied to tables, properties when applied to columns, or even sets of tables when applied at the schema level in the relational model.

\subsubsection{Display} The {\tt display} annotation is generic and configures the display name or short documentation string (i.e.\ a tool-tip) for tables and columns. This overrides values derived from the actual names and comments in the database model. In the absence of these overrides, there are also several general rules to help with display names for tables (entity types) and columns (property names). We support options to pass a model name verbatim and/or to apply transforms such as replacing underscore characters with white-space or to apply title-casing to a (presumably) lower-case table name or column name in the model.

\subsubsection{Generated} The {\tt generated} annotation is generic and emulates a read-only policy condition. Like with the {\tt required} annotation, this makes the interaction templates more strict than the underlying database platform. This can disable edits on a column or hide data-entry tools entirely for an entire table, which can be useful when a higher-privilege user is using the interface and the model contains some content that is intended to be maintained by automated pipelines and not manually entered.

\subsubsection{Immutable} The {\tt immutable} annotation is generic and emulates an insert-or-read-only policy condition. Like with the {\tt generated} annotation, it helps disable confusing or misleading interactive tools for higher-privilege users. While a generated record or column is considered to always be read-only for the interface, an immutable record or column supports data-entry for inserting new entities but disables data-entry for editing tasks on existing content.


%

%% file: parts/implementation.tex

\section{\chaise: an implementation of Model Driven Interface Generation}\label{sec:implementation}

We have implemented our model-driven interface generation approach in a browser based interaction environment called \chaise. \chaise\ is part of the \deriva\ ecosystem creating scientific asset management systems to support collaborative data-centric scientific exploration~\cite{Bugacov2017}. 

\begin{figure}[ht]
  \centering
  \includegraphics[width=1.0\linewidth]{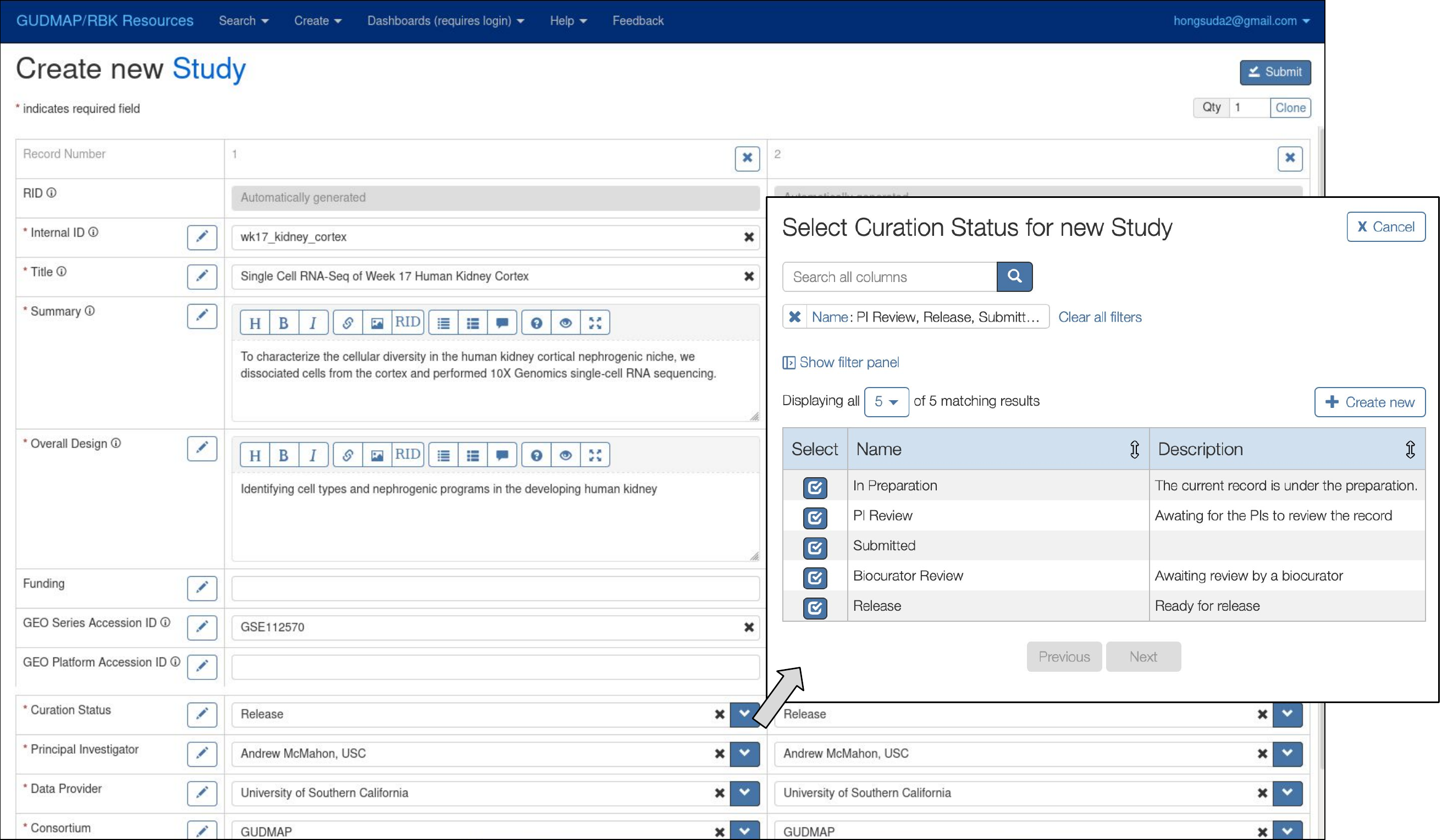}
  \caption{RecordEdit interface generated from the data-entry template for the GUDMAP/RBK deployment. A foreign key selection list is displayed for user input. Users can expand the facet panel to utilize faceted search.}
  \label{fig:gudmap-study-recordedit}
\end{figure}

\begin{figure}[ht]
  \centering
  \frame{\includegraphics[width=0.9\linewidth]{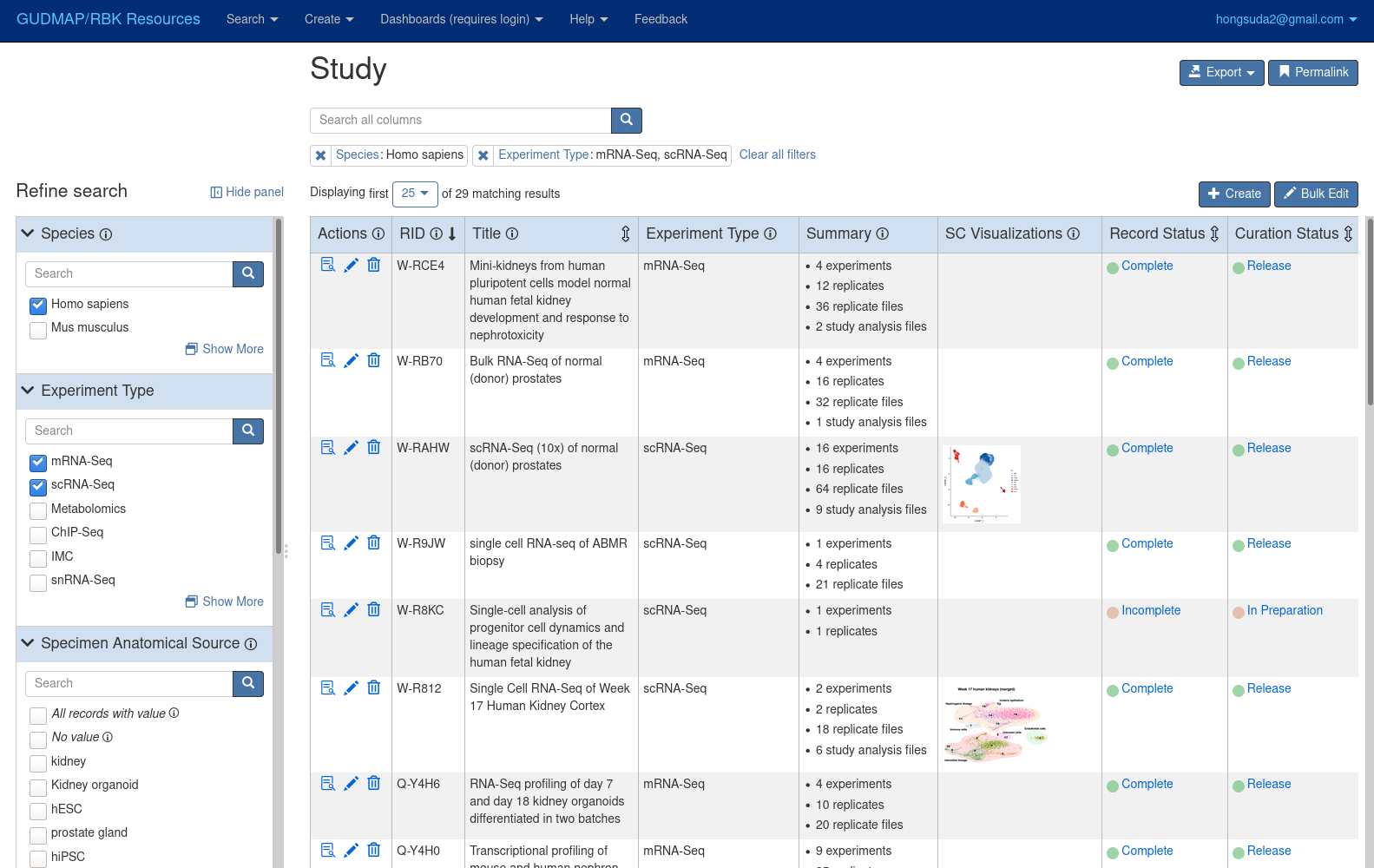}}
  \caption{RecordSet interface generated from the search template for the GUDMAP/RBK deployment}
  \label{fig:gudmap-study-recordset}
\end{figure}

\begin{figure}[ht]
\begin{subfigure}{\textwidth}
  \centering
  \frame{\includegraphics[width=0.9\linewidth]{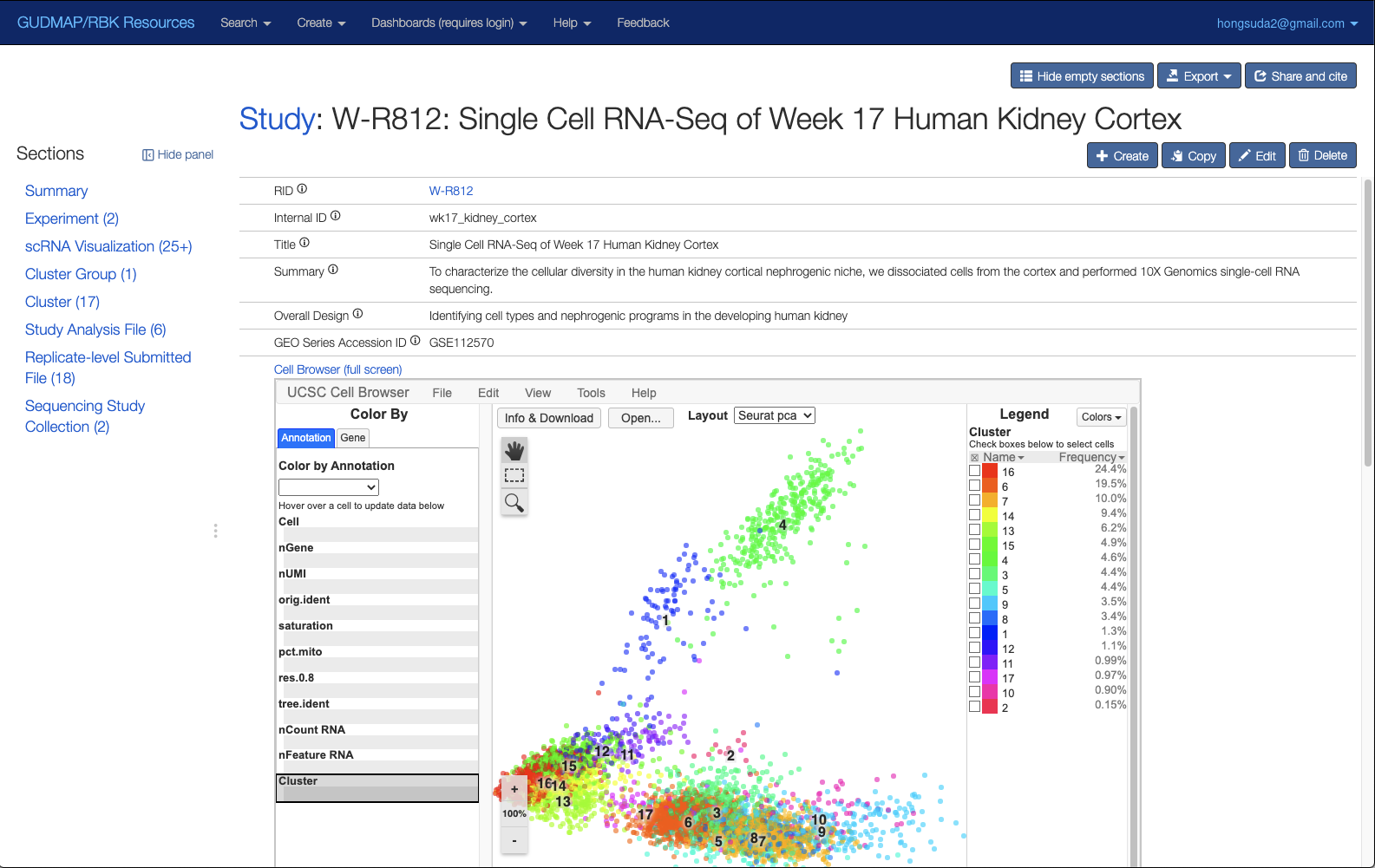}}
  \caption{Entity details}
  \label{fig:gudmap-study-record-details}
\end{subfigure}
\begin{subfigure}{\textwidth}
  \centering
  \frame{\includegraphics[width=0.9\linewidth]{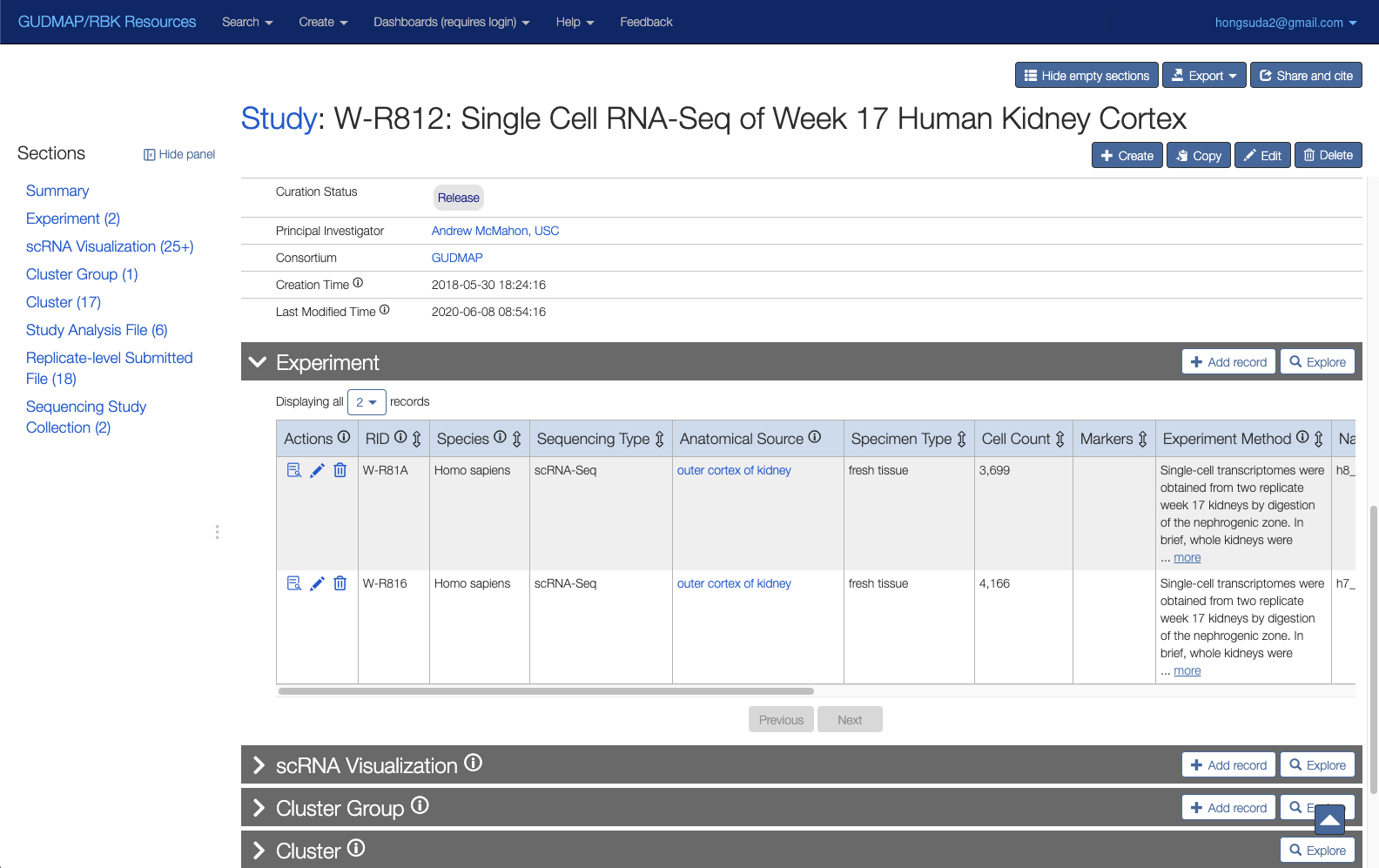}}
  \caption{Related entities}
  \label{fig:gudmap-study-record-related}
\end{subfigure}
\caption{Record interface generated from the detailed entity view template for the GUDMAP/RBK deployment}
\label{fig:gudmap-study-record}
\end{figure}

\chaise\ is implemented as a set of single-page applications---RecordEdit, RecordSet, and Record---that are developed in JavaScript using the AngularJS application framework. These applications implement the application templates described in Section~\ref{sec:apps} as a user-agent acting as a client of \deriva\ web service APIs to interact with a specialized relational data management service~\cite{Czajkowski:2018}. This data management service provides the underlying relational data storage, model management, model annotation storage, and policy enforcement mechanisms needed for our approach, as outlined in Section~\ref{sec:approach}. The storage system also provides a query interface which \chaise\ uses to extract content needed for its user interfaces.
Building on this API, our individual single-page applications all follow a common pattern to generate the concrete user interface for the interactive applications described earlier:
\begin{description}
    \item[Authentication] If necessary, the user is prompted to log in to establish the user role (client identity in Figure~\ref{fig:approach}).
    \item[Introspection] The database model is retrieved with embedded model annotations and tailored to the requesting user's access rights (the role-based model in Figure~\ref{fig:approach}).
    \item[Presentation Mapping] A model-to-presentation mapping is planned for a given application template, heuristically inferring the ER interpretation of a table (the application templates of Figure~\ref{fig:approach}).
    \item[Data Retrieval] Data is retrieved in support of the planned presentation, subject to access enforcement which may mask certain values or hide entire rows stored in the database (the role-based data of Figure~\ref{fig:approach}).
    \item[Data Presentation] The UI is rendered for the user, fully adapted to the model, presentation hints, user access rights, and retrieved content (the concrete interface of Figure~\ref{fig:approach}).
\end{description}
These common steps are explained in the remainder of this section. Each application also provides different event-driven actions for the user to perform after the initial content presentation is rendered. Figure~\ref{fig:gudmap-study-recordedit}, \ref{fig:gudmap-study-recordset}, and \ref{fig:gudmap-study-record} shows the concrete implementation of the three application templates.  

\subsection{Authorization} 
\chaise\ supports authenticated and unauthenticated users. All authentication and authorization decisions are enforced by the back end data service. \chaise\ applications facilitate login for interactive users to establish an authenticated user session, and use policy decision information (access rights summaries embedded in the role-based model) to customize the interface.  For example, edit options are not shown for tables that cannot be edited. Identity and group membership are managed by the Globus Authentication service~\cite{tuecke2016globus} which allows authentication across a wide range of identity providers and manages group membership. 

\subsection{Model Introspection} 
The ~\deriva\ platform provides a tailored model introspection response which takes into account the privilege level of the requesting user to return a detailed description of the relational model and any associated annotations. The enumerated model is further qualified with data-independent access rights of the user, i.e.\ whether the user can use the table or column for select, insert, update, and/or delete requests. For low-privilege users, parts of the model (e.g. tables, columns, and relationships) can be entirely hidden. The model description document also contains machine-readable model annotations as described in Section~\ref{sec:annotation}. To support evolution and collaboration, \chaise\ performs model introspection every time an application page loads (or reloads).   

The data-independent policy can allow a column to be hidden depending on the role of the user. In GUDMAP/RBK, only entities that are released should be visible to the public. To achieve this, a {\em Curation\_Status} column is included in all primary tables associated with different assay types, as exemplified by the Study table metadata. The {\em Curation\_Status} tracks the status of individual entities in the curation process e.g. {\em In preparation}, {\em PI review}, {\em Biocurator review}, or {\em Release}. Members of the project who manage data release can see this column and manage its content, while this concept is hidden from members of the public at large. 
The \deriva\ database platform also supports data-dependent policies to control access on a row-by-row basis. These are used to enforce row visibility rules that depend on the value in this {\em Curation\_Status} column, only showing rows to the public when the status is set to {\em Release}.

With such a policy, when an anonymous user engages with \chaise, the returned role-based model (as shown in Figure~\ref{fig:approach}) will omit all the hidden columns, and the role-based data will only include the released entities. The hidden column, and other features dependent on the hidden column, will be invisible to the public user. Of course, anonymous users have read-only access and so will see an interface optimized for searching and browsing without extraneous editing controls.

\subsection{Model-to-Presentation Mapping} 
 
Application templates use heuristics, model annotations, and built-in assignment of presentation contexts to build the most relevant ER interpretation of the database for the given user and interaction. 
In \deriva, annotation values are stored as arbitrarily-structured, machine-readable JSON documents. For annotations summarized in Section~\ref{sec:annotation}, we have defined specific JSON document structures and annotation keys which may be placed into this generic \deriva\ annotation storage, and \chaise\ validates the documents against its expectations and the discovered relational model while building this ER interpretation. Invalid parts of annotations will be pruned at narrowly-scoped boundaries, printing diagnostic information to the browser developer console but continuing with the valid subset of annotation content. This is important to provide full adaptation to the role-based model, which may prune columns, tables, or foreign keys for some users and therefore cause some custom ER property or relationship definitions to be valid for some users and invalid for others. It is also helpful for a collaboratively managed catalog, so that small configuration mistakes do not interrupt service. The downside to this approach is allowing some configuration errors to go unnoticed if the user does not recognize that the customization they tried to deploy is not actually appearing.

\deriva\ provide APIs, python/javascript libraries, and command line tools to perform Create, Read, Update, Delete (CRUD) operations on models and we have added validation rules for the annotations understood by \chaise. Users can validate annotations syntax against the models and annotation JSON schema before committing their annotation changes, to lessen the chance of deploying custom configurations which are silently ignored from the perspective of regular users.
Annotation changes require no system downtime and will be reflected during the next page load.  
Based on our science collaborators' experience, which is consistent with ours, most of the effort spent in creating annotations is not on the technical hurdle of writing an annotation, but rather on refining the custom ER abstractions and other display customizations that they wish to impose over the relational data. We have reduced the technical barrier to deploying interface changes, but exploration and user-testing~\cite{scowcroft2015exploring} are still important when trying to solve interaction problems.

\subsection{Data Retrieval, Query Planning and Execution} 
After \chaise\ builds its ER model interpretation and plans its interface layout, it 
plans queries satisfying the dynamic presentation mapping. On the back end, \deriva\ provides access aligned to the relational storage, and so these queries include the necessary table instances, filter predicates, and joins to retrieve core entity content and denormalized content. 
For example, in the experimental study model mentioned in Section~\ref{sec:requirements}, to display only human studies, the generated query involves joining the study, experiment, and biosample tables, then filter the studies to only those that contain biosamples with human species. The species concept is defined in the enriched model as a custom relationship between the study and species controlled vocabulary, chained across those other intervening tables.
The core UI elements of Record, RecordEdit, and RecordSet are data-driven pages representing one entity or entity set which is the result of a single query request. However, other surrounding UI elements and features, like reachable values in a search facet or computed aggregate properties, may involve additional queries. These extra results are progressively rendered in priority order to help mask system latency. Some UI elements act lazily, requesting and rendering data only as the user activates them.
 
\subsection{Data Presentation} 
\chaise\ implements default behaviors to render typed column values such as numbers, dates, timestamps, plain unicode text, and Markdown text without requiring any model annotations. It also implements the other heuristics and annotation-driven customization described in previous sections. As a web-based UI, \chaise\ also  allows extensive opportunities to override column or row display behaviors by specifying custom interpolation patterns. These are small template strings  expressing (conditional) interpolation of structured metadata into an intermediate Markdown fragment, which is text with some optional mark-up syntax for basic text formatting, bullet lists, embedded image tags, and links. The output of this embedded template engine~\cite{handlebars} is then fed to a Markdown renderer to produce structured HTML suitable for inclusion in the overall interface that has been generated by the application templates built into \chaise.
To enable flexible integration with other web applications and visualizations, the extended Markdown module supports inclusion of \texttt{iframe} elements. Through the {\tt iframe} feature, applications can integrate virtually any standard Web component to customize a \chaise\ deployment to its use case requirements. For example, we routinely add data visualizations such as scatter plots, histograms, as well as specialized data viewers such as high-resolution zoom-able image viewers and volume renderers.
Figure~\ref{fig:gudmap-study-record} shows an example of an embedded specialized data viewer on a Record page. Some of these extensibility features might not be feasible in an implementation of our approach which lacks the sandboxed execution and formatting environment of a modern web browser.

Finally, all the application templates produce systematic identifiers and classes in the generated HTML pages for different UI elements on the page.  This allows the styles of those elements to be overridden through the Custom Style Sheet (CSS), enabling even more avenues for interface customization without requiring redevelopment of our application templates. For example, when presenting a property containing a controlled term, a colored symbol may be added alongside the term value to visually reflect distinct states, as shown in the Record\_Status column in Figure~\ref{fig:gudmap-study-recordset}; or for a column with visualization, the column heading may be suppressed to better utilize the space, as shown by the Cell Browser in Figure~\ref{fig:gudmap-study-record}.

%% file: parts/deployments.tex
\section{Evaluation}\label{sec:deployments}

\chaise\ has been deployed to support a diversity of use cases from large centralized data hubs to small- and mid-scale collaborations including: 
\begin{itemize}
  \item The GUDMAP/RBK data repository~\cite{GUDMAP}\cite{RBK} providing curated microscope imagery, transcriptomics, cell lines, and other resources related to genito-urinary development and kidney regeneration and repair research; 
  \item The FaceBase data hub~\cite{FaceBase}, organizing a central repository for imaging and transcriptomic data generated by a number of individually-funded spoke sites; 
  \item Mapping the Dynamic Synaptome (Synapse), a multidisciplinary effort to develop methods for in vivo measurement of the synaptome; and finally 
  \item The microscopy core for the Center for Regenerative Medicine and Stem Cell Research (CIRM), offering microscope slide-scanning as a service. 
\end{itemize}
In this section, we summarize key characteristics and statistics of the above deployments to demonstrate how the aspects of our model-adaptive approach can be applied to support diverse use cases in practice.
Our evaluation includes model complexities, usage of UI customization through annotations, customization of properties and relationships, model and annotation evolution, and actual community usages of the three~\chaise\ applications.

\begin{table}[ht]
  \caption{Model complexity statistics (as of 2020/10/1)}
  \label{tab:models}
  \begin{tabular}{ l c c c c c c c c c c}
    \toprule
               &          & \multicolumn{3}{c}{\# Columns} & \multicolumn{3}{c}{\# Outbound Fkeys} & \multicolumn{3}{c}{\# Inbound FKeys} \\
    Deployment & \# Tables & Max & 95th & 50th & Max & 95th & 50th & Max & 95th & 50th \\
    \midrule
    GUDMAP/RBK & 290 & 57 & 29 &  6 & 27 & 10 &  1 & 40 & 10 &  1 \\
    Facebase   &  98 & 35 & 21 & 11 & 13 &  7 &  1 & 42 &  6 &  1 \\
    Synapse    &  34 & 33 & 31 & 10 & 11 & 11 &  2 & 13 & 11 &  1 \\
    CIRM       &  29 & 55 & 21 & 14 & 14 &  5 &  1 &  4 &  4 &  2 \\
    \bottomrule
  \end{tabular}
\end{table}

\subsection{Model Complexity}
\chaise\ is adaptive to diverse model complexity.
Table~\ref{tab:models} shows model complexity based on the number of tables, table widths (max, 95th, and 50th percentile), outbound foreign keys (max, 95th, and 50th percentile), inbound foreign keys (max, 95th, and 50th) across the 4 deployments.
The number of model-driven application instances correlates with the number of tables, but is potentially much higher, e.g.\ the Record app generates potentially unique layouts for different entities due to sparse usage of relationships, and the role-based aspects may generate different variants of the application for different users of the same tables.
Similarly, the complexity of pages correlates with the width of tables and stored foreign key relationships, but these other factors of role-based layouts and sparse data may generate a vast number of variant page layouts. 
The wider tables require more columns to be retrieved and tend to need UI customization, as not all columns should be shown in all contexts. The outbound foreign keys influence the number of joins required in a query in order to present readable display in a default interpretation of the model. The inbound foreign keys influence the number of requests needed to present the related data associated with a table of interest in the default interpretation of the model.   

\begin{figure}[ht]
\begin{subfigure}{.49\textwidth}
  \centering
  \includegraphics[width=\linewidth]{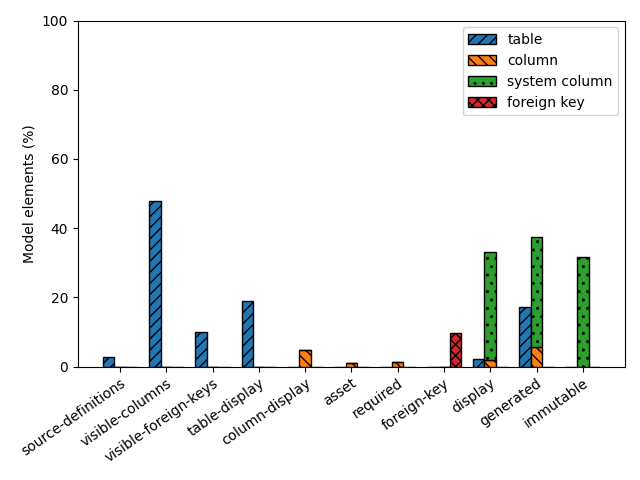}
  \caption{GUDMAP/RBK}
  \label{fig:gudmap-annotation-dist}
\end{subfigure}
\begin{subfigure}{.49\textwidth}
  \centering
  \includegraphics[width=\linewidth]{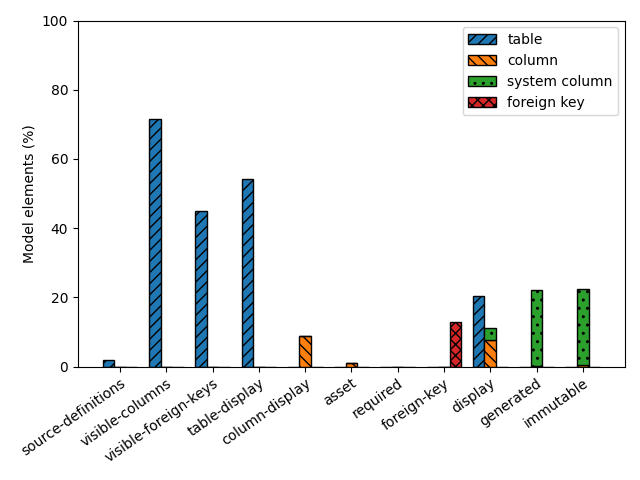}
  \caption{FaceBase}
  \label{fig:facebase-annotation-dist}
\end{subfigure}
\begin{subfigure}{.49\textwidth}
  \centering
  \includegraphics[width=\linewidth]{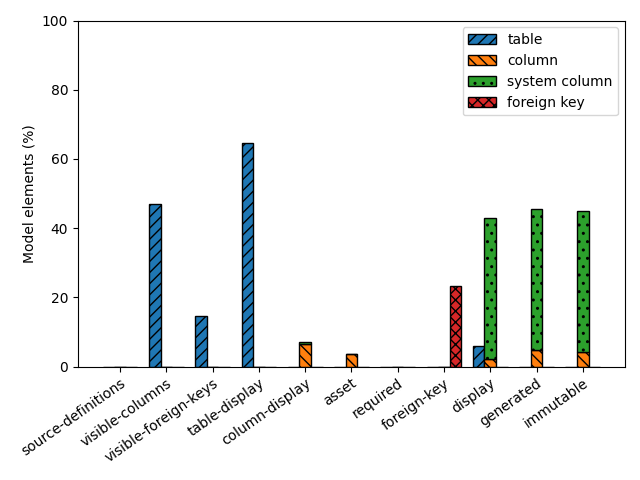}
  \caption{Synapse}
  \label{fig:synapse-annotation-dist}
\end{subfigure}
\begin{subfigure}{.49\textwidth}
  \centering
  \includegraphics[width=\linewidth]{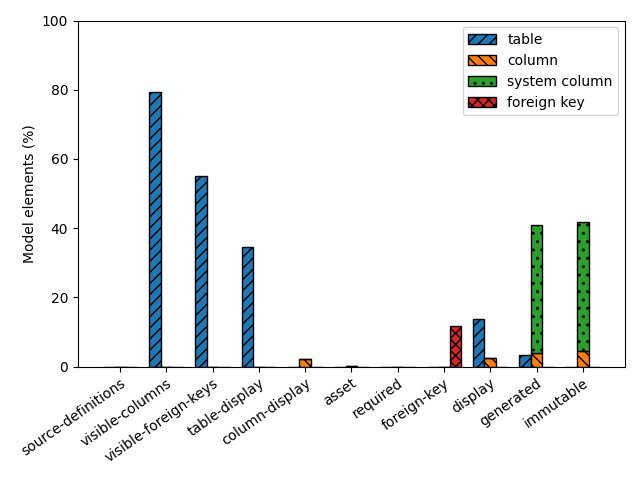}
  \caption{CIRM}
  \label{fig:cirm-annotation-dist}
\end{subfigure}
\caption{Annotation distribution}
\label{fig:annotatio-distributions}
\end{figure}

\subsection{UI Customization}
\chaise\ supports multiple customization options for its interfaces. 
Figure~\ref{fig:annotatio-distributions} shows the normalized annotation distribution across different annotation keys (Table~\ref{tab:annotations}) used to configure these options. 
Many annotation keys are targeted toward specific model classes, while some can be applied to multiple classes. The labels/colors in the figure correspond to different model classes that each annotation key is applied. For column-level annotations, the {\em column} and {\em system column} labels capture annotations applied to user-defined columns and {\em system columns}, respectively. The system columns---namely, {\tt RID}, {\tt RCB}, {\tt RMB}, {\tt RCT}, {\tt RMT} representing row ID, row creation and modified by, and row creation and modified time, respectively---were introduced in later \deriva\ release as a requirement in individual tables to enable data versioning and provenance.   
The annotation count is normalized by dividing the number of annotation occurrences by the total number of potential sites associated with each model class (e.g. a normalized table annotation is the number of annotated tables divided by the total number of tables). 

All deployments utilize various types of annotations to address their different use cases.  
Across all deployments, the table-level annotations are extensively utilized, while the column and foreign key annotations are sparsely used. The top table-level annotations are {\tt visible-columns}, {\tt visible-foreign-keys}, and {\tt table-display}. The visible-columns are mostly used to suppress and reorder columns in different contexts, especially when there are large number of columns. For example, a small subset of columns are shown in the compact form for a quick summary, only columns relevant to user inputs are shown for editing to avoid visual clutter, and (almost) all columns are shown for a record-centric view. 
The most often use column-level annotations are {\tt generated}, {\tt immutable}, and {\tt column-display}, most of which are applied to system columns. These annotations can be removed in the future when \chaise\ expands its support to allow system column configuration at the catalog level. 

Note that a large number of annotations do not necessarily imply bad~\chaise\ heuristics, as many annotations only provide extra hints to modulate heuristics e.g. disable user input. In addition, annotations can be dormant and get applied only when they become relevant (e.g. column annotations are only applied when the columns are visible). We observe quite a few dormant annotations in some deployments (e.g. GUDMAP/RBK mark all {\tt ID} column generated). Moreover, some annotations are legacy and no longer needed as \chaise\ expands its functionality e.g. no need to specify visible columns just to suppress or rearrange system columns, or custom presentation of URL columns as links when also annotating them as assets which enjoy the built-in support for file upload and download.  
\chaise\ will continue to evolve to provide more functionality and reduce the annotation burden to support diverse deployment use cases. 


\subsection{Properties and Relationship Customization}
In this section, we evaluate the customization of columns and relationships explicitly annotated in {\tt visible-columns} and {\tt visible-foreign-keys} which are directly responsible for the page content shown on the three templates. Though these are two separate annotations, \chaise\ effectively combines them for the {\em detailed} presentation context in the Record application, so we also combine them in this analysis.

For each annotated element, we classify its definition into 4 different categories: simple column, simple relationship, complex column, and complex relationship. A {\em simple column} is a native column in the table of interest. A {\em simple relationship} include a direct relationship to its immediate neighbors or a relationship through a binary association, which are things that \chaise\ can present automatically via its heuristic interpretation of the database. A {\em complex column} is an aggregate column or a single- or multi-relationship path that leads to a column. A {\em complex relationship} is multi-relationship path that results in a set of entities. The usage of simple definitions suggests a scenario where the customization may only be reordering or hiding content that the default behavior would support. The usage of complex definitions indicates that the customization is actually imposing novel ER concepts which \chaise\ would never infer on its own.     

\begin{figure}[ht]
\begin{subfigure}{.49\textwidth}
  \centering
  \includegraphics[width=\linewidth]{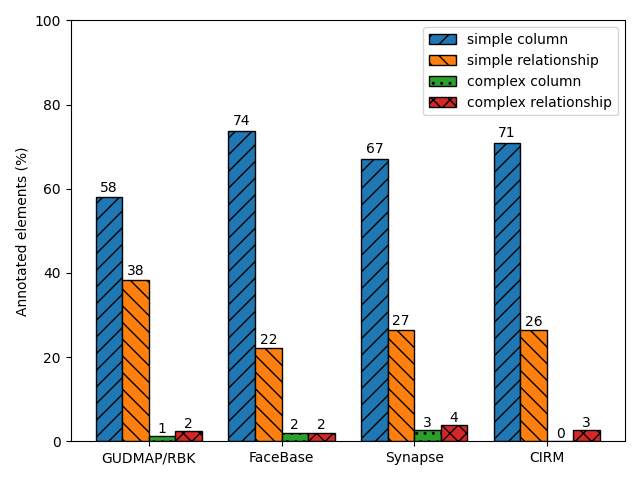}
  \caption{Annotated elements}
  \label{fig:denormalization-flat}
\end{subfigure}
\begin{subfigure}{.49\textwidth}
  \centering
  \includegraphics[width=\linewidth]{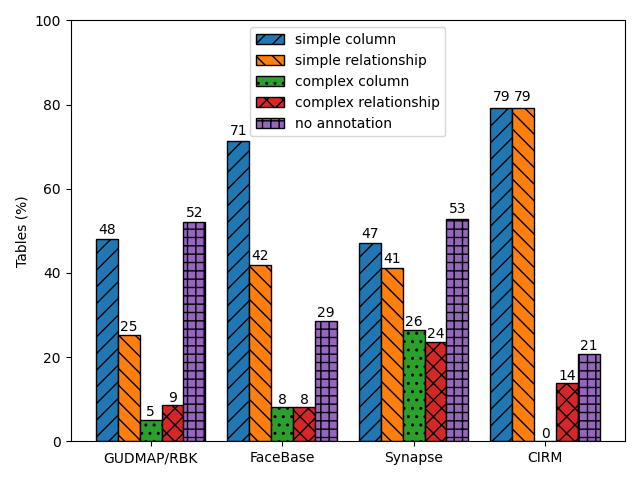}
  \caption{Tables}
  \label{fig:denormalization-table}
\end{subfigure}
\caption{Distribution of different types of columns and relationships}
\label{fig:denormalization}
\end{figure}

Figure~\ref{fig:denormalization} shows the distribution of different types of columns and relationships. All deployments utilize complex columns and relationships sporadically. We found that in the hub deployments, the complex columns are mostly used to show aggregate data in the compact context, while in Synapse, they are used for searching on columns in the filter context. There is no usage of complex columns in CIRM. Across all deployments, the complex relationships are primarily utilized in the filter context to improve search-ability, follow by visible-foreign-keys for showing multi-hops related tables. Figure~\ref{fig:denormalization-table} shows the percentage of tables with annotations containing at least one annotated element categorized as each of the categories defined above. The low fractions of tables with complex columns or relationships implies that the visible-columns and visible-foreign-keys are mostly used to suppress or re-arrange columns and relationships. Smaller deployments seem to have higher fractions of tables annotated with complex columns and relationships.   

\subsection{Model and Annotation Evolution}

\begin{figure}[ht]
\begin{subfigure}{.49\textwidth}
  \centering
  \includegraphics[width=\linewidth]{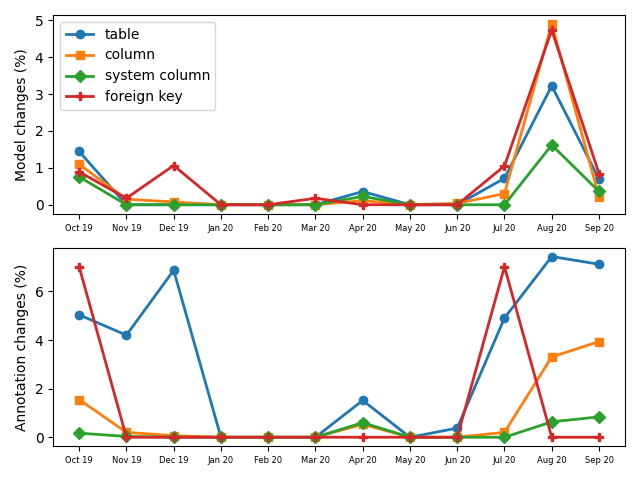}
  \caption{GUDMAP/RBK}
  \label{fig:gudmap-monthly-changes}
\end{subfigure}
\begin{subfigure}{.49\textwidth}
  \centering
  \includegraphics[width=\linewidth]{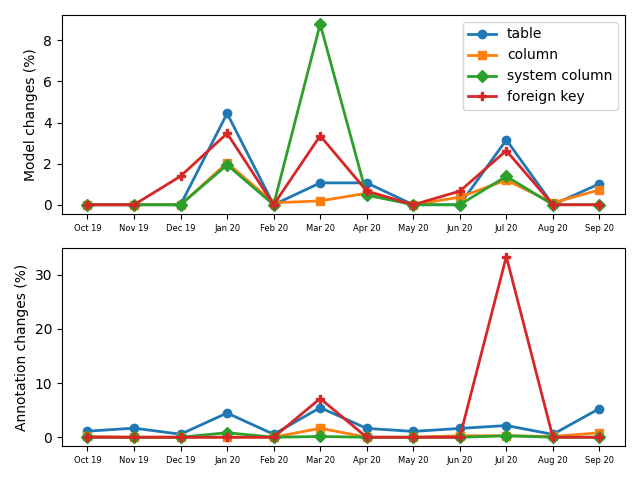}
  \caption{FaceBase}
  \label{fig:facebase-monthly-changes}
\end{subfigure}
\begin{subfigure}{.49\textwidth}
  \centering
  \includegraphics[width=\linewidth]{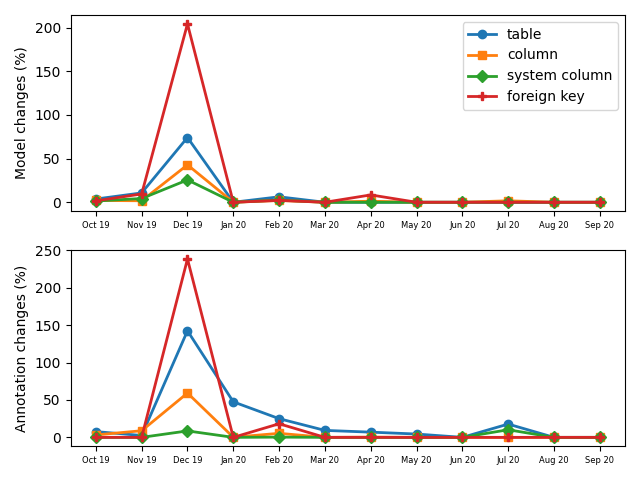}
  \caption{Synapse}
  \label{fig:snapse-monthly-changes}
\end{subfigure}
\begin{subfigure}{.49\textwidth}
  \centering
  \includegraphics[width=\linewidth]{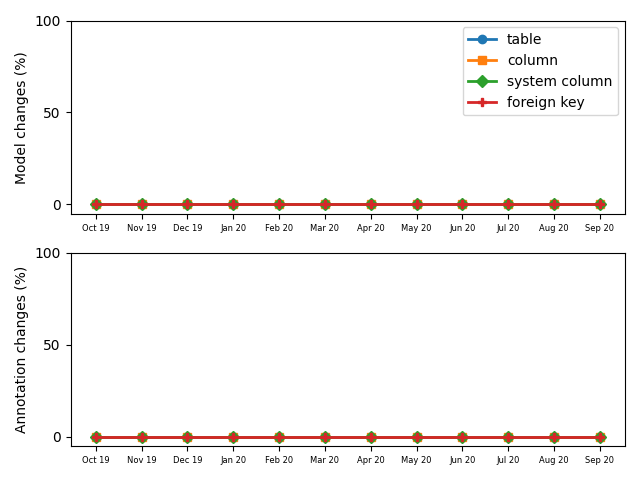}
  \caption{CIRM}
  \label{fig:cirm-monthly-changes}
\end{subfigure}
\caption{Monthly model and annotation changes from 2019/10/01 to 2020/09/30}
\label{fig:monthly-changes}
\end{figure}

\chaise\ supports model and annotation changes as deployments evolve. Figure~\ref{fig:monthly-changes} displays the monthly changes of models and annotations across all deployments from Oct 1, 2019, to Sep 30, 2020. The model and annotation changes associated with each model class are normalized by the total number of model elements and the annotations, respectively, within each class reported in the previous month. The {\em column} and {\em system-column} are user-defined columns and system-specific columns introduced earlier. 

There were multiple model and annotation changes made in all deployments except CIRM over the last 12 months. As expected, the changes are bursty---there were a few larger changes mixed in with no or small changes. The bigger changes correspond to model extensions to support new assay types or use cases. For example, the model changes in GUDMAP/RBK in Sep 2020 were due to the image annotation addition. Similarly, the spike in changes for Synapse reflects a shift from a completely private collaboration to one supporting a mixture of unreleased and public results. Across projects, the smaller changes usually relate to model tweaking to add or remove metadata fields, types, or linkage. The annotation changes associated with model changes are usually deployed together. However, we also see annotation changes that happened independently of model changes. This is to be expected as presentations can continue to be tweaked for better user experiences or to refactor the annotations when \chaise\ released new features. 
\chaise\ has been successfully used to support model and annotation changes as the science investigations evolve to accommodate more use cases since their inception circa 2016.

\subsection{Users and Data Usage}

\chaise\ supports diverse user communities and their usage of scientific data. 
Table~\ref{tab:data} summarizes user and data usage statistics of four different user communities through the use of the three application templates between a six-month period from May 1, 2019, to Oct 31, 2019. 
The table shows: 1) the number of users representing the scale of user communities; 2) the number of data requests across three different \chaise\ templates and data mutation requests representing the data usage; and 3) the table coverage---the number of tables involved in all \chaise\ requests---representing the model exposure.  
Traffic from within our organization is excluded. The number of \chaise\ requests only reflects the main page events; subsequent requests within the same pages are excluded. Individual mutation requests can involve one or multiple rows. 

\begin{table}[ht]
  \caption{Deployment statistics over a 12 month period from 2019/10/01 to 2020/09/30}
  \label{tab:data}  
  \begin{tabular}{l c c c c c c c}
    \toprule
    & & \multicolumn{4}{c}{\# Data Requests} & Data & Table \\
    Deployment & \# Users & Recordset & Record & Recordedit & Total & Mutation  & usage \\ 
    \midrule
    GUDMAP/RBK & 6,064 & 15,956 & 33,218 & 5,285 & 54,459 & 5,340 & 75\% \\
    FaceBase   & 8,268 &  9,646 & 18,006 & 2,157 & 29,809 & 2,279 & 88\% \\
    Synapse    &   426 &  3,468 &  8,939 & 2,523 & 14,930 & 2,576 & 97\% \\
    CIRM       &    76 &     33 &     43 &   21  &     97 &    23 & 66\% \\
    \bottomrule
  \end{tabular}
\end{table}

The data hub deployments (GUDMAP/RBK, FaceBase) have bigger user communities and hence more data requests compared to the smaller closed collaboration deployments (Synapse, CIRM). GUDMAP/RBK and FaceBase are similarly used to curate and publish final data products. There are more assay types in GUDMAP/RBK compared to FaceBase, which potentially explains the higher mutation requests. 
In Synapse, \chaise\ plays a key role in facilitating day-to-day collaboration activities to conduct complicated science experiments. Therefore, it is not surprising to see high data usage despite its smaller user community.    
CIRM data is primarily acquired through an automatic ingest pipeline and \chaise\ is mostly used to view and/or edit data; this practice is reflected in the low number of mutation requests. Note that the CIRM activities of this period are only 20\% comparing to the 12-month period (2019/03/01-2020/02/29) leading to the lab shutdown in March 2020. The model exposure among all deployments is relatively high. The unexposed tables are primarily legacy tables that are no longer linked to the primary model or data navigation menus.     
Our deployment data demonstrates \chaise's adaptability to different user sizes, communities, and usage patterns.




%% file: parts/related-work.tex
\section{Related Work}\label{sec:related}

The challenges of handling increasing volumes of data in the discovery process have been approached from several different but related angles. Storage management systems, such as the Storage Resource Broker (SRB)~\cite{Baru1998a}, Integrated Rule-Oriented Data Service (iRODS)~\cite{Rajasekar2010a}, and Storage Resource Manager (SRM)~\cite{Shoshani2002a}, are primarily focused on data archival and transfer services with minimal support for descriptive metadata in a predominately key-value structure. While key-value stores provide some utility for basic ``tagging'' with simple descriptive terms, scientific workloads have been shown to require more complex data representation and querying~\cite{Jain2016}, thus motivating the need for rich meta-models such as the ER model supported in our approach. More specialized scientific data management platforms such as the Open Microscopy Environment Research Objects (OMERO)~\cite{Swedlow2009a} and the eXtensible Neuroimaging Archive Toolkit (XNAT)~\cite{Marcus2007a} go a step beyond the generic data management of SRB, iRODS, and SRM by adding more complex scientific metadata handling that users can model through JavaScript Object Notation (JSON) or eXtensible Markup Language (XML) templates, respectively. While these allow for greater description of the scientific domain than the key-value data stores, they still fall short of the richness of relational and ER models to represent complex relationships between many conceptual entities that comprise a given scientific domain of inquiry. Though these approaches do allow the user to customize the representation of the domain captured in the metadata, our approach allows for arbitrarily complex metadata models and dynamically generates the interfaces on the fly without the need for offline configuration and rebuild of the user interfaces.

Specialized metadata catalog services have been explored as a means of complementing storage management systems with additional capabilities for the description of scientific data. These catalogs provide interfaces for scientists to interact with research datasets for the purpose of discovery, organization, and publication with the need for model evolution specifically called out~\cite{Deelman2004a}. The data modeling capabilities vary widely across different metadata catalogs from key-value stores similar to the storage management approaches to somewhat more expressive Entity-Attribute-Value (EAV)~\cite{Koblitz2007a} and also XML~\cite{Jensen2006a} based models. While these approaches have utility for managing basic descriptions of datasets, they are again limited compared to the complexity possible in ER models as required by the workloads observed in scientific discovery. In addition, unlike our approach to fully generating a usable interaction environment, these metadata catalogs support only generic user interfaces that do not adapt tailored interfaces to the domain model.


Among enterprise information systems, database workbenches are commonly available for many open source and commercial database management systems, such as PostgreSQL Admin Tools~\cite{Page2020a} or SQuirreL~\cite{Bell2020a} for instance. These workbenches support a wide range of tasks including database server administration, data model design and implementation, backup and restore, system logging and reports, interactive querying, and data manipulation. While it could be argued that these workbenches provide interfaces for search and manipulation of data and they clearly adapt to changing schema of the database, they lack a task specification, do not simplify the rendering of potentially complex data structures, and provide little assistance for generating non-trivial queries needed for data discovery. Beyond these workbenches, numerous database user interface development environments exist some with simplified visual development interfaces, such as the Oracle Visual Builder~\cite{Oracle2020a}. These and other similar environments reduce the development effort to build a database-oriented application, however, they still require development skills typically beyond the training of our target users (scientists and data scientists), and unlike our approach, they do not generate applications that dynamically evolve with the model in real-time.

The broader category of Model-Based User Interface Development (MBUID)~\cite{Meixner2011a} have informed our approach. We follow the pattern of having a separation of concerns based on a set of declarative models~\cite{Da-Silva2000a} for the UI generation: the domain model is specified in the design of the relational storage model, the presentation model is specified by annotating the relational model with hints to infer a more abstract ER model, and the user model is expressed in the form of declarative, role-based access control statements also embedded within the model. However, our approach centers all modeling aspects on the extended model thereby keeping a tight cohesion across these interrelated concerns and maintaining its focus on the data-driven workloads of the generated interfaces. Also, in our approach, the application templates (Section~\ref{sec:apps}) serve as a form of pre-defined Abstract User Interfaces (AUIs)~\cite{Engel2013a}, and therefore alleviate unnecessary burden from the system modeler while still addressing the stereotyped interactions of search, view, and edit as needed by our target use cases. Our approach constructs the concrete interfaces from these application templates through the heuristics and annotations described in Section~\ref{fig:approach}.

Automated user interface development~\cite{Martin2013a} could greatly improve the viability of producing interactive applications for data-driven science by drastically reducing the burden for developing and maintaining user interfaces over the lifetime of a scientific investigation. 
Existing techniques, however, often progress through stages of only partially automated user interface generation interspersed with dependent activities for business analysts, system modelers, and user interface developers. In order for the benefit of automated UI generation to be fully realized in science, the UI generation must be fully automated and minimize the burden on the scientist to develop elaborate model specifications that evolve over the course of an investigation. Though our approach accepts annotations on the relational model to influence the automated, heuristic-based UI generation, these annotations are optional. A system modeler can begin without annotations and slowly influence the UI generation process over time as requirements and expertise in applying annotations grow. A pattern library as proposed in PaMGIS~\cite{Engel2009a} could, however, enrich our approach beyond the fixed set of application templates to enable specialized tasks beyond our core set for search, display, and edit of structured information.


We describe our approach as being \textit{model-adaptive} as one of our underlying motivations is that models used in data-driven discovery change frequently over the course of an investigation. Database schema evolution is notoriously challenging in large part because of the impact it has on database-oriented applications that are typically hard-wired to a particular database schema unless applications can evolve with the database~\cite{Stonebraker2016a}. Our approach allows the applications to adapt in real-time to automatically reflect changes to the underlying model. Alternative approaches to \textit{adaptive} user interface generation, such as Role-Based UI Simplification (RBUIS)~\cite{Akiki2013a}, adapt the user interface to different user roles in environments where the enterprise applications, such as Customer Relationship Management (CRM) or Enterprise Resource Planning (ERP), have expansive feature sets and not all features are relevant to all users of the system. Such adaptation tailors the UI to a particular class of users according to the functions of greatest interest to them. These approaches do not, however, address the issues of model-adaptation required to keep a database-oriented UI in sync with the evolution of the database schema.

MBUID approaches have been proposed for generation of database-oriented applications. 
An early example, GENIUS~\cite{Janssen1993a}, generated user interface specifications from the database model and introduced ``Dialog Nets'' based on Petri Nets to formalize task modeling. Teallach~\cite{Griffiths2001a}, an MB-UID environment for object databases, introspects the underlying database to generate the domain model similar to our approach, but still requires user intervention to define the task model and presentation models.
DB-USE~\cite{Tran2010a} proposes a process for UI generation for database-oriented applications from a set of domain, task, and user models with analysts, designers, and developers playing interrelated roles throughout the process. Like our approach, DB-USE automatically generates not only the UI but also the database query and manipulation functions necessary to support the user interactions. Unlike DB-USE, Teallach, and GENIUS, our approach goes a step further to not only infer the domain model from the database but also to heuristically-generate UI applications from a set of application templates. A key benefit of our approach is that it can quickly adapt to changes in the underlying database schema.

%% file: parts/conclusion.tex
\section{Conclusion}\label{sec:conclusion}

In this paper, we have summarized the key requirements to the use of database technology in collaborative scientific investigations. We propose a model-adaptive approach to user interface generation, describe its implementation in \chaise---a suite of dynamic, model-driven, web applications for curating and discovering scientific data organized within relational databases. 
Our solution relies on a networked, relational database storage system to provide a collaboratively-maintained metadata repository (addressing Requirement~\ref{sec:req-collaborate}).
Our model-adaptive approach automatically generates interactive GUIs for editing, searching, and viewing scientific data based on the introspection of a relational data model comprising aspects of domain, user, and presentation models (addressing Requirement~\ref{sec:req-web-uis}). 
The model-driven approach enables our UIs to support diverse and evolving data models and access policies (addressing Requirement~\ref{sec:req-diverse-models}). Our application templates abstract the foreign keys in a relational database as relationships for richer search and presentation by automatically displaying related content as part of a more intuitive entity-relationship presentation. Data modelers can customize this ER interpretation of their model (addressing Requirement~\ref{sec:req-fk-paths}). Finally, our approach supports customization and extension of data presentation through annotations to address other deployment specific needs (addressing Requirement~\ref{sec:req-annotation}).  

\chaise\ has been deployed for actual use in research projects ranging from small laboratories, to core facilities, to large research consortia across highly diverse and evolving domains. 
We present the key deployment characteristics and statistics to demonstrate \chaise\ adaptability and flexibility.  
As part of our future work, we plan to develop UI-based annotation editor to help guide users through annotation configuration; simplify the configuration of system-columns en masse; further optimize the query generator; and improve the data curation flow to make it more user friendly. \chaise\ is available as part of the open source \deriva\ platform for scientific data management and is in daily use by active scientific collaborations.

